\documentclass[aps,prb,twocolumn,showpacs,superscriptaddress]{revtex4-1}  % for review and submission

\usepackage{graphicx}  % needed for figures
\usepackage{dcolumn}   % needed for some tables
\usepackage{bm}        % for math
\usepackage{amssymb}   % for math
\usepackage{amsmath,epsfig}   % e.g. for boxes
\usepackage{verbatim}  % to use \begin{comment}
\usepackage{cases}
\usepackage{color}

\newcommand{\fgies}[3]{\mbox{\raisebox{#3}
{\epsfig{file=#1,scale=#2,clip=true}}~}}
\newcommand{\bml}{\begin{multline}}
\newcommand{\bea}{\begin{eqnarray}}
\newcommand{\eea}{\end{eqnarray}}
\newcommand{\be}{\begin{equation}}
\newcommand{\ee}{\end{equation}}
\newcommand{\bi}{\begin{itemize}}
\newcommand{\ei}{\end{itemize}}

\newcommand{\rr}{\mathbf{r}}
\newcommand{\kk}{{\mathbf{k}}}
\newcommand{\pp}{\mathbf{p}}
\newcommand{\qq}{\mathbf{q}}

\newcommand{\PP}{\mathbf{P}}
\newcommand{\vn}{\mathbf{0}}

\newcommand{\ktyp}{k_{\rm typ}}

\newcommand{\up}{\uparrow}
\newcommand{\down}{\downarrow}

\newcommand{\Ar}{\mathcal{A}}
\newcommand{\Br}{\mathcal{B}}

\newcommand{\rP}{\mathcal{P}}
\newcommand{\Cr}{\mathcal{C}}
\newcommand{\Tr}{\mathcal{T}}
\newcommand{\Zr}{\mathcal{Z}}
\newcommand{\Dr}{\mathcal{D}}
\newcommand{\Sr}{\mathcal{S}}
\newcommand{\Qr}{\mathcal{Q}}
\newcommand{\Nr}{\mathcal{N}}

\newcommand{\ddelta}{\boldsymbol{\delta}}
\newcommand{\kkappa}{\boldsymbol{\kappa}}
\newcommand{\Gz}{G^{(0)}}
\newcommand{\td}{\tilde}

\newcommand{\bcbp}{\begin{comment}}
\newcommand{\ecbp}{\end{comment}}

\newcommand{\bl}{} %\color{blue}}

\begin{document}

\title{\bl 
Diagrammatic Monte Carlo algorithm for the resonant Fermi gas}

\author{K. Van Houcke}
\affiliation{Laboratoire de Physique 
de l'Ecole Normale Sup\'erieure, ENS - PSL, CNRS, Sorbonne Universit\'e,
Universit\'e Paris-Diderot, Sorbonne Paris Cit\'e, 75005 Paris, France}
\affiliation{Department of Physics, University of Massachusetts, Amherst, MA 01003, USA}
\author{F. Werner}
\affiliation{Laboratoire Kastler Brossel, Ecole Normale Sup\'erieure - PSL, Sorbonne Universit\'e, Coll\`ege de France, CNRS, 24 rue Lhomond, 75005 Paris, France}
\affiliation{Department of Physics, University of Massachusetts, Amherst, MA 01003, USA}
\author{\bl T. Ohgoe}
\affiliation{Department of Applied Physics, University of Tokyo,
7-3-1 Hongo, Bunkyo-ku, Tokyo 113-8656, Japan}
\author{N. V. Prokof'ev}
\affiliation{Department of Physics, University of Massachusetts, Amherst, MA 01003, USA}
\affiliation{Russian Research Center ``Kurchatov Institute", 123182 Moscow, Russia}
\author{B. V. Svistunov}
\affiliation{Department of Physics, University of Massachusetts, Amherst, MA 01003, USA}
\affiliation{Russian Research Center ``Kurchatov Institute", 123182 Moscow, Russia}
\affiliation{Wilczek Quantum Center, School of Physics and Astronomy and T. D. Lee Institute, Shanghai Jiao Tong University, Shanghai 200240, China}

\date{\today}

\begin{abstract}
We provide a description of 
a diagrammatic Monte Carlo algorithm for the
resonant Fermi gas in the normal phase.
Details are given 
on diagrammatic framework, Monte Carlo moves, and incorporation of ultraviolet asymptotics.
{\bl Apart from the self-consistent bold scheme, we also describe a non-self-consistent scheme, for which the ultraviolet treatment is  more involved.}
\end{abstract}

%\pacs{}
\maketitle

\section{Introduction}

A major long-standing challenge is to find a method for solving a 
generic fermionic many-body problem in the thermodynamic limit with controlled accuracy.
The diagrammatic technique is the most versatile quantum-field-theoretical tool allowing one to
express the answers as series of integrals of a special structure. Each term in the series can be 
visualized with graphs---Feynman diagrams---built using simple rules.
In  the absence of small parameters, there is little hope to sum the 
diagrammatic series analytically, and one commonly resorts to uncontrollable truncations.
In contrast, the goal of
the Diagrammatic Monte Carlo (DiagMC) approach is to sum up all Feynman diagrams in a systematic way up to a controlled accuracy.
{\bl Using an efficient Monte Carlo algorithm to evaluate
all diagrams up to a high enough order $N_{\rm max}$,
convergence as a function of $N_{\rm max}$ can be observed,
as first demonstrated for the Hubbard model~\cite{VanHoucke1,KozikVanHouckeEPL}.
The thermodynamic limit is taken from the outset since one works only with connected diagrams.
Furthermore one can build diagrams with fully-dressed propagators;
this self-consistent formulation, called Bold Diagrammatic Monte Carlo (BDMC),
was first demonstrated for the Fermi-polaron~\cite{ProkofevSvistunovPolaronLong} and the resonant Fermi gas~\cite{VanHouckeEOS}.
}

The resonant  Fermi gas is a 
{\bl three-dimensional continuous-space model}
 of great interdisciplinary interest. 
It features a smooth crossover between fermionic and bosonic superfluidity,
as argued in the context of condensed matter physics\cite{Leggett1980_1,Leggett1980_2,NSR,ChapLeggettBref}
and later observed experimentally in ultracold atomic Fermi gases near Feshbach resonances.\cite{ZwergerBook}
The model is also relevant to neutron matter\cite{GezerlisNeutronsReview} and high-energy physics,\cite{NJP_QCD}
in particular in the unitary limit where the scattering length is infinite.

{\bl
For the unitary Fermi gas in the normal unpolarized phase,
first BDMC results
for the equation of state  were reported in Ref.~\onlinecite{VanHouckeEOS}.
Very recently, 
these results were confirmed using
a much
more
advanced resummation method,
which was found to be necessary for controllability,
due to the fact that the series has zero radius of convergence~\cite{RossiEOS}.
Contact and momentum distribution were also computed using the new resummation method~\cite{RossiContact}.
In the meantime, 
the DiagMC approach was also developed further and applied to frustrated quantum magnetism~\cite{KulaginPRB,KulaginPRL,HuangPyro}
and 
various lattice models of interacting
fermions~\cite{GukelbergerPwave,DengEmergentBCS,SimonsHubbardBenchmark,GukelbergerFFLO,IgorDirac,kozik_pseudogap,GukelbergerIsingAF,JohanInfiniteU,GullDualFermions,KozikDualFermions,SimonsHydrogenChain}
including models with electron-phonon interaction~\cite{MishchenkoProkofevPRL2014,IgorCoulombPhonon}
and topological phase transitions~\cite{IgorHaldane}.

In this paper,
we describe
the numerical method
used for the equilibrium
normal resonant Fermi gas
in Refs.~\onlinecite{VanHouckeEOS,RossiEOS,RossiContact}, in particular how to evaluate the terms of  the diagrammatic series to high orders (typically up to order 9) using a diagrammatic Monte Carlo algorithm, and how to properly incorporate large-momentum asymptotics coming from the contact interactions.
Another crucial ingredient is the proper resummation of the divergent diagrammatic series~\cite{RossiEOS}
which will be detailed
elsewhere~\cite{ResonLong2}.}

{\bl We mostly use a bold diagrammatic scheme,
where diagrams are} built with fully dressed single-particle propagators and pair propagators.
We present a set of elementary Monte Carlo updates 
to sample this diagrammatic space.
While 
some features of
the updating scheme are analogous to the ones introduced for the bare series of the Hubbard model in Ref.~\onlinecite{VanHoucke1},
an important difference 
is that only fully irreducible skeleton diagrams are sampled,
so that ergodicity has to be carefully verified.
Furthermore,
resonant fermions
feature 
specific ultraviolet singularities
governed by an observable called contact.\cite{TanEnergetics,TanLargeMomentum,ChapLeggettBref,ChapBraatenBref,LeChapitreBref}
This physics manifests itself in a natural way within our skeleton diagrammatic framework,
and is readily incorporated into our BDMC scheme.
{\bl For cross-validation, we use not only the self-consistent bold scheme, but also a non self-consistent ``ladder scheme'', in which case the ultraviolet physics governed by the contact can also be incorporated semi-analytically, using a more elaborate procedure.}

The paper is organized as follows.
In Section~\ref{sec:diag}, the diagrammatic framework is constructed,
arriving at the skeleton series for the single-particle  and pair self-energies. %in terms of the fully dressed single-particle propagator and pair propagator.
Section~\ref{sec:algo}
describes
the diagrammatic Monte Carlo algorithm:
The diagrammatic expansion
is expressed as a Monte Carlo average in subsection~\ref{subsec:general_idea},
precise descriptions of
configuration space, probability density
and measurement procedure
are given in subsections~\ref{subsec:config}, \ref{subsec:weight} and \ref{subsec:measuring},
the update scheme
is described in subsection~\ref{subsec:updates},
 reducibility and ergodicity issues are discussed in subsection~\ref{subsec:ergodic},
{\bl
the self-consistent iteration procedure is described in subsection~\ref{subsec:iter},
and resummation is briefly
mentioned in
Sec.~\ref{sec:resum}.}
Section~\ref{sec:UV} describes ultraviolet analytics and its incorporation into BDMC.
{\bl The ladder scheme is treated in Sec.~\ref{sec:ladder}.}

\section{Diagrammatic framework}
\label{sec:diag}

\subsection{The resonant Fermi gas model}

In the zero-range model, also known as the resonant gas model,
the interaction is characterized by
the $s$-wave scattering length $a$.
The zero-range model is a universal limit of finite-range models.
More precisely,
a generic interaction of range $b$ can be replaced by the zero-range model
in the 
limit where $b$ becomes much smaller than other typical lengthscales of the problem, 
such as the interparticle distance,
the thermal wavelength, and $|a|$.
For an  atomic alkali Fermi gas near a broad Feshbach resonance,
the range is set by the van der Waals length, 
and most current experiments are well within the zero-range limit, with finite-range corrections in the percent or sub-percent range.\cite{WernerCastinRelationsFermions}

Even though our Monte Carlo scheme works directly with the zero-range interaction in continuous space,
it is convenient to start with a lattice model, thereby eliminating ultraviolet divergences at the initial steps of constructing the formalism.
The Hamiltonian reads
\bml
\hat{H'}  =   \hat{H}-\sum_ {\sigma= \uparrow, \downarrow}\mu_{\sigma} \hat{N}_\sigma   =   \sum_{\sigma= \uparrow, \downarrow}   \sum_{\mathbf{k}\in\Br}\,~\left(
\epsilon_\kk-\mu_{\sigma}\right)\,\hat{c}_{\kk,\sigma}^\dagger \hat{c}^{\phantom{\dagger}}_{\kk,\sigma} \\
 + g_{0} \, b^3  \sum_{\rr}  ~
\hat{n}_\uparrow(\mathbf{r})
\hat{n}_\downarrow(\mathbf{r}) \, ,\qquad \qquad  \qquad \qquad \qquad
\label{eq:hamil}
\end{multline}
where the spin index $\sigma$ takes on the values $\up$ and $\down$,
the operator $ \hat{c}^{\phantom{\dagger}}_{\kk,\sigma}$
 annihilates a spin-$\sigma$ fermion of momentum~$\kk$,
$\hat{\psi}^{\phantom{\dagger}}_{\sigma}(\mathbf{r})$  is the
 corresponding position-space annihilation operator,
$\hat{n}_\sigma(\rr)=\hat{\psi}^\dagger_\sigma(\rr)\hat{\psi}_\sigma(\rr)$
is the number-density operator,
  $\mu_{\sigma}$ is the spin-dependent chemical potential,
  $\hat{N}_{\sigma}$ is the number operator for spin-$\sigma$ fermions,
   $\rr$ is a position vector whose components are integer multiples of
the lattice spacing   $b$  (this $b$ can also be viewed as the interaction range since the interaction is on-site),
$\Br = ]-\pi/b, \pi/b]$ is the first Brillouin zone,
and the dispersion relation is
$\epsilon_\kk=k^2/2$ with particle mass set to unity.~\footnote{As usual, it is implicit that, in the sum over momentum in Eq.~(\ref{eq:hamil}),  formally considering a finite system before eventually taking the thermodynamic limit, the coordinates of $\kk$ are integer multiples of $2\pi/L$ where $L$ is the length of the cubic box with periodic boundary conditions. Also, for the lattice model, $\epsilon_\kk=k^2/2 $ holds only
for $\kk\in\Br $, while $\epsilon_\kk$ is extended outside of $\Br$ by periodicity.}
The bare coupling constant $g_0$ is adjusted to have the desired scattering length $a$ for two particles on the lattice in free space,
namely
\begin{equation}
\frac{1}{g_0} = \frac{1}{4\pi a} - \int_\Br \frac{d\kk}{(2\pi)^3} \frac{1}{k^2} \;.
\label{eq:scatlen}
\end{equation}
%where $m$ is twice the reduced mass,  $2/m \equiv 1/m_\up + 1/m_\down$.
The zero-range limit  corresponds to the continuum limit $b{\to}0$, with $a$ fixed.
One can note that $g_0\to0^-$ in this limit. %, as follows from Eq.~(\ref{eq:scatlen}).

\subsection{Single-particle propagator, self-energy, and ladder summation} %$G$ and $\Sigma$}

In the standard diagrammatic formalism for the many-body problem at finite temperature,\cite{FetterWalecka,PitaevskiiLifchitz,AGD}
the central object is the single-particle propagator %-- also called Green's function,
\be
G_\sigma(\pp,\tau) = -\left< {\rm T}\ \hat{c}^{\phantom{\dagger}}_{\pp,\sigma}(\tau)\hat{c}^\dagger_{\pp,\sigma}(0)\right> \; ,
\label{eq:onegreen}
\ee
where $\tau$ is the imaginary time and ${\rm T}[\ldots]$ is the time-ordered product.
This Green's function gives access to
the momentum distribution
$n_\sigma(\pp)= G_\sigma(\pp,\tau=0^-)$,
and 
to the number density~\footnote{We use the following standard convention for the Fourier transformation between position and momentum space:
$f(\rr) = \int f(\pp) e^{i \kk \cdot \rr} \, d\pp/(2\pi)^3$.
It is implicit that the integrals over momenta run over the first Brillouin zone $\Br$ when working with the lattice model, and over the entire space $\mathbb{R}^3$ when working with the zero-range model in continuous space.}
\be
n_\sigma=G_\sigma(\rr=\vn,\tau=0^-).
\label{eq:n_vs_G}
\ee
In the series expansion of $G$ in powers of the bare coupling constant $g_0$, each term can be represented by a Feynman graph:
\be
\fgies{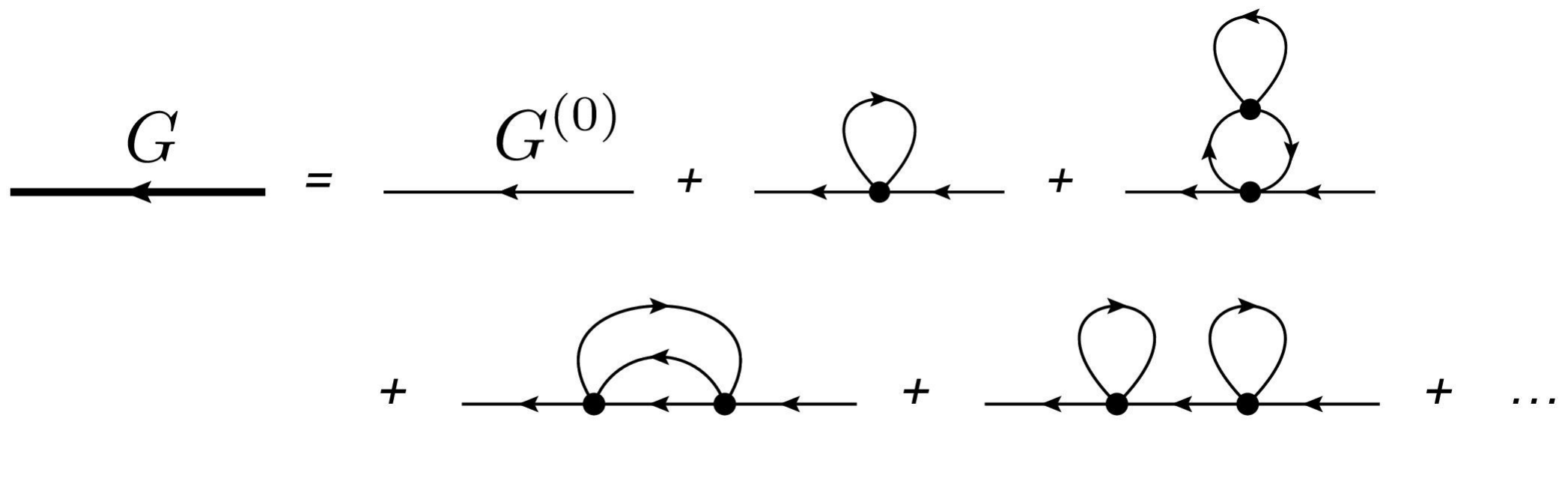}{0.25}{0mm}
\ee
where the bare interaction vertex $\bullet$ denotes $g_0$,
the thin lines denote an ideal gas propagator $G^{(0)}$,
and the bold line denotes the fully dressed (i.e. exact) propagator $G$.

The first natural step to organize the higher-order terms is to introduce the self-energy $\Sigma$, which is related to $G$ by the
Dyson equation,
given diagrammatically by
\be
\fgies{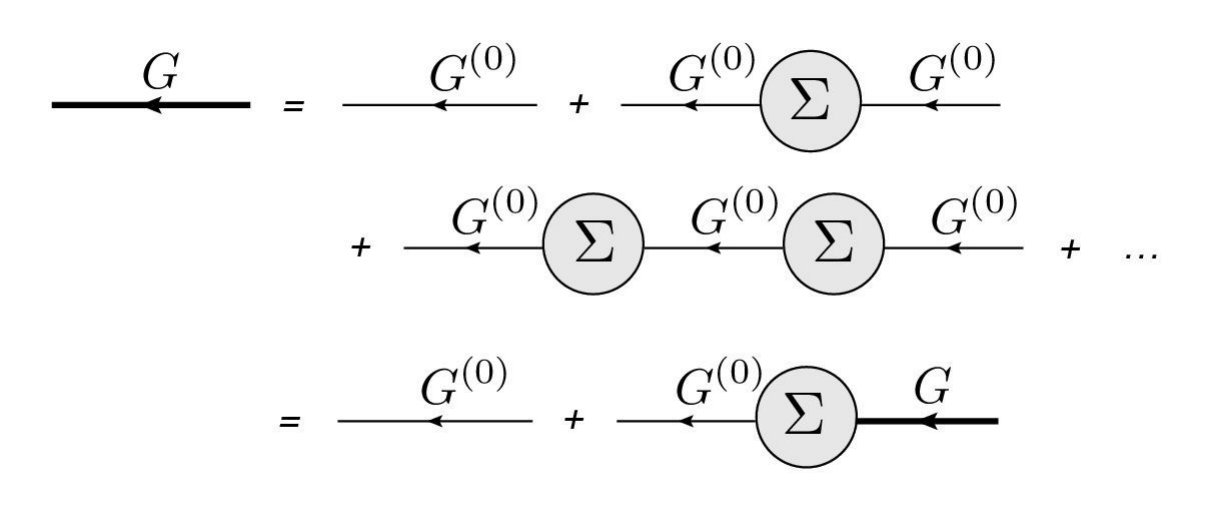}{0.4}{0mm}
\label{eq:dyson}
\ee
i.e.
\be
\frac{1}{G_\sigma(\pp,\omega_n)} = \frac{1}{G^{(0)}_\sigma(\pp,\omega_n)}- \Sigma_\sigma(\pp,\omega_n)
\label{eq:dyson_G}
\ee
for any fixed momentum $\pp$ and Matsubara frequency $\omega_n$. \footnote{We use the following standard notations for the Fourier transformation between imaginary time  and Matsubara frequencies: for a $\beta$-antiperiodic function $f(\tau)$,
$f(\omega_n) = \int_0^\beta d\tau f(\tau) e^{ i \omega_n \tau}$ where
 $\omega_n=(2n+1)\pi/\beta$ are the fermionic Matsubara frequencies;
while for a $\beta$-periodic function $f(\tau)$,
$f(\Omega_n) = \int_0^\beta d\tau f(\tau) e^{ i \Omega_n \tau}$ where
 $\Omega_n=2n\pi/\beta$ are the bosonic Matsubara frequencies.}
To avoid double counting, reducible diagrams are excluded from
$\Sigma$, so that
\be
\fgies{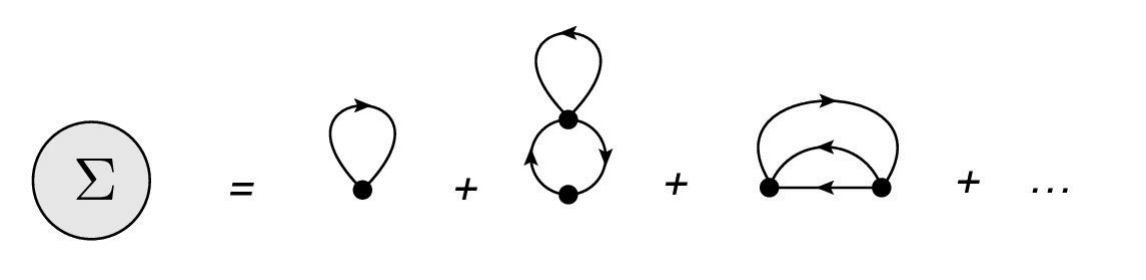}{0.4}{0mm}
\label{eq:Sigma_g0}
\ee

Another standard step is to perform  summation of ladder diagrams:
\be
\fgies{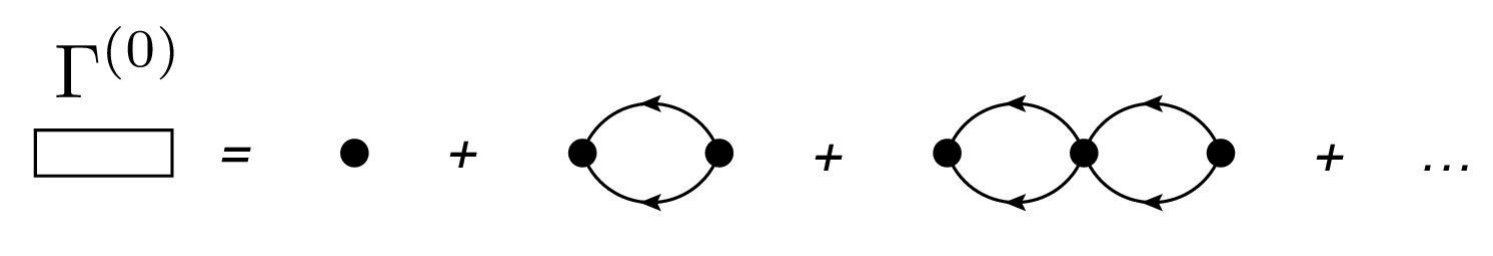}{0.3}{0mm}
\label{eq:Gamma0_diag}
\ee
Physically, such a ladder summation is natural since in vacuum it would correspond to the two-body scattering amplitude or $T$-matrix.
This allows one to take the zero-range limit and work directly with zero-range interactions in continuous space.
$\Gamma^{(0)}$ is an approximate pair propagator, which can also be viewed as a renormalized interaction vertex;
eventually, $\Gamma^{(0)}$ will be replaced by
a fully dressed pair propagator in
our BDMC scheme.
Summation of the geometric series in Eq.~(\ref{eq:Gamma0_diag}) gives
\be
\frac{1}{\Gamma^{(0)}(\mathbf{P},\Omega_n)} =   \frac{1}{g_0}     - \Pi^{(0)}(\mathbf{P},\Omega_n)
\label{eq:Gamma0_Pi0}
\ee
where
$\Pi^{(0)}$ is the $(G^{(0)}\,G^{(0)})$ bubble
given by
\bea
\Pi^{(0)}(\mathbf{P},\Omega_n)   =     - \beta^{-1} \sum_m  \int_\Br \frac{d\mathbf{k}}{(2\pi)^{3}}  ~G^{(0)}_{\uparrow} (\mathbf{P}/2+\mathbf{k}, \omega_m) \nonumber \\
 ~~\cdot ~G^{(0)}_{\downarrow} (\mathbf{P}/2- \mathbf{k}, \Omega_n - \omega_m)  \nonumber \\
  =     \int_\Br \frac{d\mathbf{k}}{(2\pi)^{3}}   \frac{1 - n^{(0)}_{\uparrow}(\mathbf{P}/2 + \mathbf{k}) - n^{(0)}_{\downarrow}(\mathbf{P}/2 - \mathbf{k}) }{i \Omega_n +
2\mu
- \epsilon_{\PP/2 + \kk}
- \epsilon_{\PP/2 - \kk} }\; ,
\nonumber
\eea
with the Fermi factor
$n_{\sigma}^{(0)}(\mathbf{k})=[1 + e^{\beta (\mathbf{k}^2/2 -\mu_{\sigma})}]^{-1}$.
The integral over $\kk$ is finite thanks to the restriction to the first Brillouin zone $\Br$.
Here $\beta$ is the inverse temperature and $\mu = (\mu_\up+\mu_\down)/2$ is the mean chemical potential.
 Eliminating the bare coupling constant $g_0$ in Eq.~(\ref{eq:Gamma0_Pi0}) in favor of the scattering length $a$---using relation (\ref{eq:scatlen})---finally yields
\begin{multline}
\frac{1}{\Gamma^{(0)} (\mathbf{P},\Omega_n)}  = \frac{1}{4\pi a} -  \int \frac{d\mathbf{k}}{(2\pi)^{3}}
 \Bigg[ \frac{1}{k^2}
\\+ \frac{1 - n^{(0)}_{\uparrow}(\mathbf{P}/2 + \mathbf{k}) - n^{(0)}_{\downarrow}(\mathbf{P}/2 - \mathbf{k}) }{i \Omega_n +
2\mu - P^2 /4 - k^2  }
\Bigg] \; ,
\label{eq:gammagen}
\end{multline}
where the integration domain for $\kk$ is now taken to be $\mathbb{R}^3$ instead of $\Br$, i.e.  the continuum limit is taken.
The diagrammatic expansion of the self-energy can then be written in terms of the vertex $\Gamma^{(0)}$ instead of $g_0$; to avoid double counting one simply has to forbid diagrams containing $(G^{(0)}\,G^{(0)})$ bubbles:
\be
\fgies{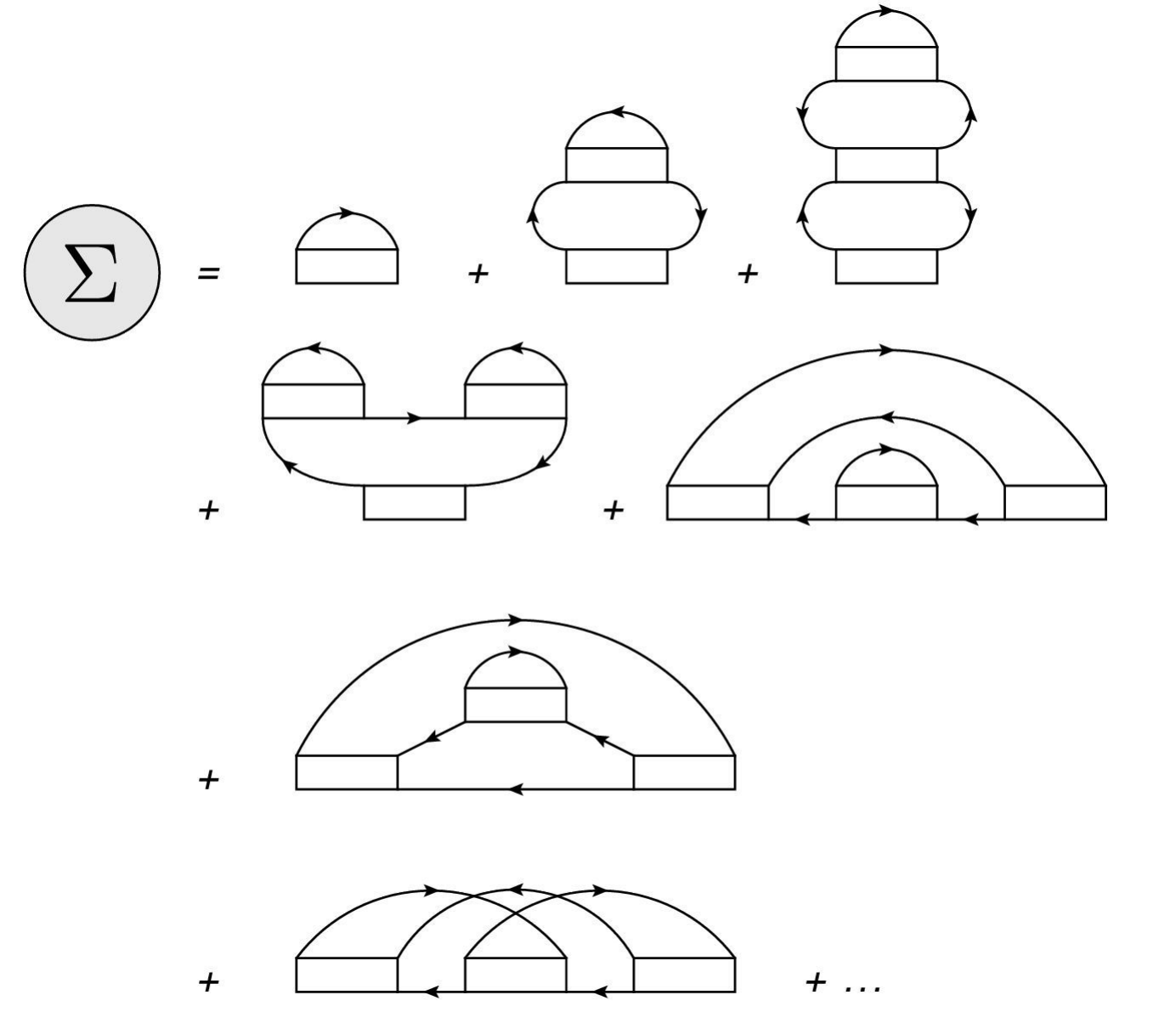}{0.3}{-1mm}
\label{eq:Sig_vs_Gamma0}
\ee
{\bl
Here each $G^{(0)}$ line is meant to have a fixed spin label, which is not shown for simplicity.
}
We thus arrive at the exact diagrammatic representation of the zero-range {\bl continuous-space} model to be used in what follows.

Many diagrammatic studies of the BEC-BCS crossover problem are based on the bare $T$-matrix, $\Gamma^{(0)}$, and the lowest-order diagram for $\Sigma$ in terms of $G^{(0)}$ and $\Gamma^{(0)}$, see, e.g., Refs.~\onlinecite{NSR,ChapStrinatiBref,combescot_leyronas_kagan}.
For example, this approximation is sufficient for obtaining the exponential scaling of the critical temperature $T_c \propto e^{-\pi/(2k_F|a|)}$ in the BCS limit.

%%%%%%%%%%%%%%%%%%%%%%%%%
\subsection{Bold pair propagator}

While the diagrammatic elements introduced in the previous section are completely standard,
a more original aspect of our diagrammatic framework is the use of a fully dressed (bold) pair propagator $\Gamma$.
In the case of the polaron problem, this was done in Refs.~\onlinecite{ProkofevSvistunovPolaronLong,Vlietinck13}.
The propagator $\Gamma$ is defined by
\be
\Gamma(\pp,\tau)=g_0\,\delta(\tau)+g_0^{\phantom{0}2}\cdot\rP(\pp,\tau)
\label{eq:GammavsP}
\ee
with
\be
\rP(\rr,\tau)\equiv-\left< {\rm T}\ (\hat{\Psi}_\down \hat{\Psi}_\up)(\rr,\tau)(\hat{\Psi}^\dagger_\up\hat{\Psi}^\dagger_\down)(\vn,0)\right>\; ,
\ee
or, diagramatically,
\be
\fgies{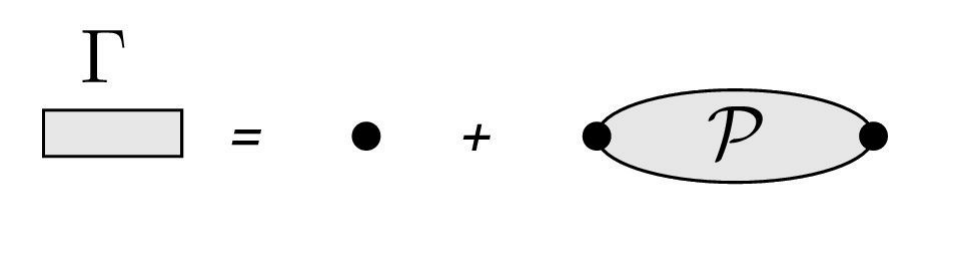}{0.3}{0mm}
\label{eq:GammavsP_diag}
\ee
One can note that the first term in Eqs.~(\ref{eq:GammavsP},\ref{eq:GammavsP_diag}) goes to zero
in the continuum limit.

Similarly to the Dyson equation that expresses the bold single-particle propagator $G$ in terms of the irreducible single-particle self-energy $\Sigma$
[Eq.~(\ref{eq:dyson})],
we can write a Dyson equation for the bold pair propagator $\Gamma$ in terms of an irreducible pair self-energy $\Pi$:
\be
\fgies{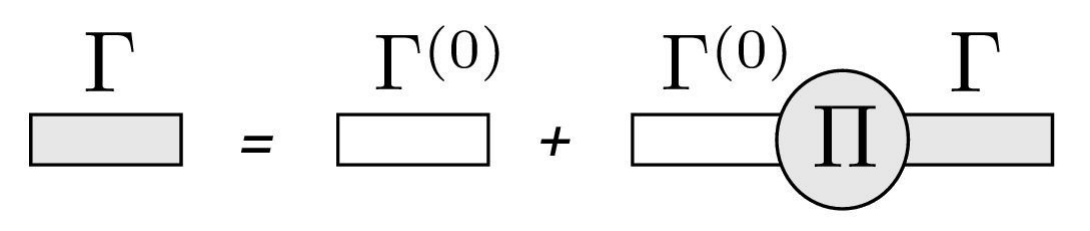}{0.3}{-2mm}
\ee
i.e.
\be
\frac{1}{\Gamma(p,\Omega_n)} = \frac{1}{\Gamma^{(0)}(p,\Omega_n)}-\Pi(p,\Omega_n).
\label{eq:dyson_Gamma}
\ee

\subsection{Feynman rules for the skeleton diagrams}\label{subsec:feyn_rules}

Bold diagrammatic Monte Carlo
works with skeleton diagrams
built with fully dressed (bold) lines.
For the unitary Fermi gas, we use diagrams built from the bold single-particle propagator $G_{\sigma}$  and the bold pair propagator $\Gamma$ defined above. 
The first diagrams expressing the single-particle self-energy $\Sigma$ in terms of $G$ and $\Gamma$ are
\be
\fgies{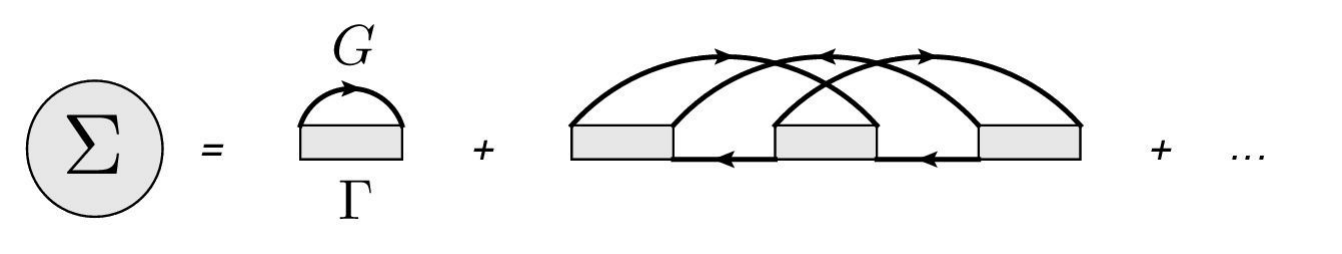}{0.4}{0mm}
\label{eq:Sig_skele}
\ee
while the first diagrams for the pair  self-energy $\Pi$ are
\be
\fgies{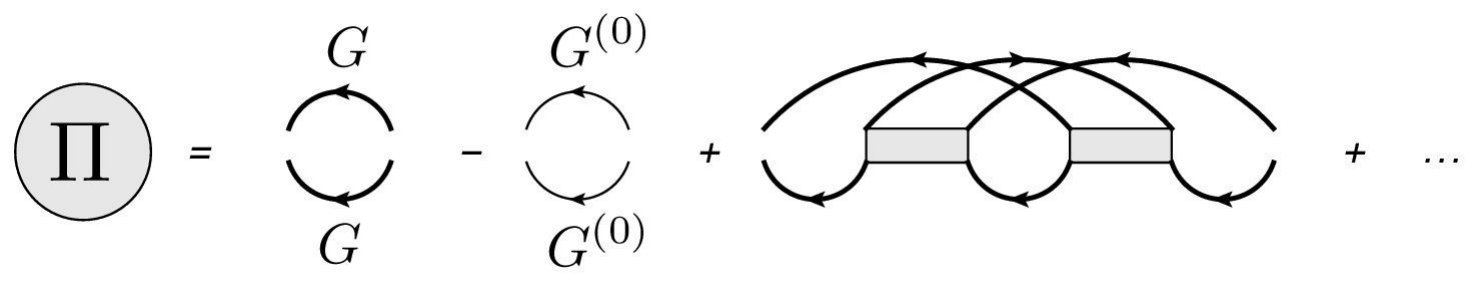}{0.35}{0mm}
\label{eq:Pi_skele}
\ee
In summary, the propagators $G$ and $\Gamma$ are expressed in terms of the self-energies $\Sigma$ and $\Pi$ through the Dyson equations (\ref{eq:dyson_G},\ref{eq:dyson_Gamma}), and the self-energies are themselves expressed in terms of the propagators through the diagrammatic expansions ~(\ref{eq:Sig_skele},\ref{eq:Pi_skele}).

Since the Feynman rules for these diagrammatic expansions are the ones which our algorithm has to obey, we describe them in some detail.
{\bl The goal is to express the sum
$\Sigma_\sigma^{{\bl(N)}}$ or $\Pi^{\bl(N)}$
 of all order-$N$ skeleton diagrams.}
We define the order $N$ of a skeleton diagram through the number of $\Gamma$-lines:
a $\Sigma$-diagram contains $N$  such lines
and a $\Pi$-diagram contains $N-1$ such lines.
Let us use the notation ${\mathcal Q}$ to denote either $\Sigma_\sigma^{{\bl(N)}}$  or $\Pi^{\bl(N)}$,
and let
$\Sr_\Qr$ be the set of all skeleton diagram topologies for $\Qr$, meaning that all these diagrams are irreducible with respect to cutting any two internal lines of the same type ({\it i.e.} the diagram should remain connected  if one cuts two $G_{\sigma}$ lines or two $\Gamma$ lines).
We shall use the shorthand notation $Y=(\pp;\tau_1,\tau_2)$ for the external diagram variables.
Clearly,
$\Qr(Y)=
\Qr(p,\tau_1-\tau_2)$. 

For a given topology, we can label each internal line by an index $l$ for a $G$-line (resp. $\lambda$ for a $\Gamma$-line), and
denote the corresponding internal momentum by $\kk_l$  (resp. $\kkappa_\lambda$), and the spins of $G$-lines by $\sigma_l$. Similarly,
the time-differences between the end and origin points of the lines are denoted
by $\Delta\tau_l$ (resp. $\Delta\tau'_\lambda$).
It can be shown that for any topology $\Tr$ in a diagram of order $N$, one can always
find $N$ `loop momenta' $\qq_1,\ldots,\qq_N$ that, together with the external momentum,
uniquely determine all the internal momenta.
More precisely, some of the internal momenta are equal to a loop momentum, while the others
are linear combinations of loop momenta such that momentum is conserved at each vertex.\footnote{The well-known fact that one can always choose $N$ independent momenta in this way can be proven rigorously by using a covering tree of the diagram (J.~Magnen, {\it private communication}).}
For our Feynman diagrams, the internal variables $X$ can thus be parameterized by $\qq_1,\ldots,\qq_N$, as well as by the internal times $\tau_3,\ldots,\tau_{2N}$  which belong to $[0,\beta]$
(these times are assigned to three-point vertices which connect a $\Gamma$-line with two $G$-lines).
With these notations, the {\bl $N$-th order of the} diagrammatic expansion simply reads
\be
\Qr(Y) = \sum_{\Tr\in\Sr_\Qr} \int dX\,\Dr(\Tr,X,Y)
\label{eq:Q_feyn}
\ee
with the differential measure
\be
dX \equiv d\qq_1\ldots d\qq_N\,d\tau_3\ldots d\tau_{2N} \;
\label{eq:dX}
\ee
and
\bml
\Dr(\Tr,X,Y) = \frac{(-1)^N\,(-1)^{N_{\rm loop}}}{(2\pi)^{3N}} \,
\\
 \times \left[\prod_l G_{\sigma_l}(\kk_l,\Delta\tau_l)\right]
\times
\,\left[\prod_\lambda\Gamma(\kkappa_\lambda,\Delta\tau_\lambda')\right] \; ,
\end{multline}
with $N_{\rm loop}$ the number of closed fermion loops in the diagram of topology $\Tr$.
{\bl There is one exception: To avoid double counting, in the first-order diagram for $\Pi$, we have to compensate for the fact that all $(G^{(0)}G^{(0)})$ bubbles are already contained in $\Gamma^{(0)}$:
\be
\Pi^{(1)}(p,\tau) = - \frac{1}{(2\pi)^3}\,\int d\qq\
\Dr
\ee
with
\be
\Dr =
G_\up(\qq,\tau)\,G_\down(\pp-\qq,\tau)
\\
-
\Gz_\up(\qq,\tau)\,\Gz_\down(\pp-\qq,\tau)\;.
\label{eq:subG0G0}
\ee
Note also that 
a diagram topology $\Tr$ is defined here by a graph with fixed spin labels.}

Note that
if we restrict to the lowest order diagram in Eq.~(\ref{eq:Sig_skele}) and~(\ref{eq:Pi_skele}),
our framework becomes
 equivalent to
 the approach introduced in Refs.~\onlinecite{Haussmann_Z_Phys,Haussmann_PRB}.
This approach is called self-consistent $T$-matrix approximation, because
$\Gamma$ is then given by the ladder diagrams built with $G$.

\section{Diagrammatic Monte Carlo Algorithm} \label{sec:algo}

\subsection{From diagrams to Monte Carlo} \label{subsec:general_idea}

In this section, we explain how the diagrammatic expansion of the previous section can be formally rewritten as a stochastic average.
As in
Refs.~\onlinecite{ProkSvistFrohlichPolaron,ProkofevSvistunovPolaronLong,VanHoucke1},
the general idea is that
the integral over internal variables $X$ and the sum over topologies $\Tr$
 will be evaluated stochastically,
 for all values of the external variables $Y$,
 through a single Monte Carlo process.
Specifically, 
 in order to determine the function $\Qr(Y)$, where $\Qr$ stands as above for $\Sigma_\sigma^{{\bl (N)}}$ or $\Pi^{\bl{(N)}}$,
we shall compute overlaps of the form
\be
\Ar_{\Qr,g} \equiv \int dY \,\Qr(Y) g(Y)
\label{eq:Q_g}
\ee
for a set of functions $g$ given below.
Expanding $\Qr(Y)$ in terms of Feynman diagrams as in Eq.~(\ref{eq:Q_feyn}) yields
\be
\Ar_{\Qr,g} =\sum_{\Tr \in \Sr_\Qr} \int dX dY \,\Dr(\Tr,X,Y) g(Y)\; .
\label{eq:feyn}
\ee
Defining a configuration by
\be
\Cr=(\Tr,X,Y),
\ee
i.e. by a given topology and given values of internal and external variables,
the expression (\ref{eq:feyn}) can be rewritten as a weighted average over configurations,
\be
\Ar_{\Qr,g} = \int d\Cr\   |\Dr(\Cr)| \cdot  {\rm sgn}[\Dr(\Cr)]\cdot  g(Y)\cdot 1_{\Tr\in\Sr_\Qr} \; .
\label{eq:int_dC}
\ee
Here we introduced the indicator function
\begin{equation*}
    1_{\Tr\in\Sr_\Qr} =
    \begin{cases}
     1, & {\rm if\ } \Tr\in\Sr_\Qr \; , \\
     0, & {\rm otherwise} \;  ,  \\
    \end{cases}
\end{equation*}
 so that the integral over $\Cr$ can be extended to topologies outside of $\Sr_\Qr$.
Our choice of the extended space of configurations will be discussed below.

In order to evaluate (\ref{eq:int_dC}) by Monte Carlo, it should be rewritten in the form 
\be
\Ar_{\Qr,g} =  \int d\Cr\ w(\Cr)\,A_{\Qr,g}(\Cr) \; ,
\label{eq:expect_val}
\ee
where $w(\Cr)\geq0$ and the total weight
\be
\Zr \equiv \int d\Cr w(\Cr)
 \label{eq:w_Z}
\ee
is finite so that
 $w(\Cr)/\Zr$ is a  normalized probability distribution.
In practice we take
\be
w(\Cr)=|\Dr(\Cr)|\,R(\Cr) \; ,
\label{eq:reweight}
\ee
where $R(\Cr)$ is an  arbitrary (non-negative) reweighing function.
It is then clear that Eq.~(\ref{eq:int_dC}) can indeed be rewritten as Eq.~(\ref{eq:expect_val}) provided we set
\be
A_{\Qr,g}(\Cr) =  \frac{{\rm sgn}[\Dr(\Cr)] \cdot g(Y) \cdot1_{\Tr\in\Sr_\Qr}}{R(\Cr)}\; .
\label{eq:estim_A_C}
\ee
The Monte Carlo update scheme (described in Sec.~\ref{subsec:updates}) will generate a Markov chain of random configurations $\Cr_1,\Cr_2,\ldots$ with the stationary probability distribution $w(\Cr)/\Zr$.
The average over $n$ generated configurations then converges to the true expectation value in the large $n$ limit,
\be
\Ar_{\Qr,g} = \Zr \times \, \lim_{n\to\infty}
\frac{1}{n}\sum_{i=1}^n A_{\Qr,g}(\Cr_i) \; .
\label{eq:A_Z}
\ee

It remains to estimate $\Zr$, which can be done easily in the following way.
 The trick
is  to have a subset $\Sr_\Nr$ of the configuration space, which we call the normalization-sector, whose total weight
\be
\int_{\Sr_\Nr}\,d\Cr\,w(\Cr) =: \Zr_\Nr \; ,
\ee
is easy to calculate analytically.
In our case, we artificially create this normalization sector by enlarging the configuration space, as we shall see below (in contrast, in the case of the Hubbard model algorithm of Ref.~\onlinecite{VanHoucke1}, the normalization sector was that of the first-order diagram).
Defining the ``norm" $\Nr$ as the number of times that the normalization sector was visited,
\be
\Nr\equiv\sum_{i=1}^n 1_{\Cr_i\in\Sr_\Nr} \; ,
\ee
$\Zr$ can be evaluated  thanks to
$\Zr_\Nr/\Zr = \lim_{n\to\infty}\Nr/n$. Inserting this into Eq.~(\ref{eq:A_Z}) yields the final expression
\be
\Ar_{\Qr,g} = \Zr_\Nr\lim_{n\to\infty}\frac{\sum_{i=1}^n A_{\Qr,g}(\Cr_i)}{\Nr} \; .
\label{eq:estim_A}
\ee

\subsection{Configuration space}\label{subsec:config}

\begin{figure}
\includegraphics[width=\columnwidth]{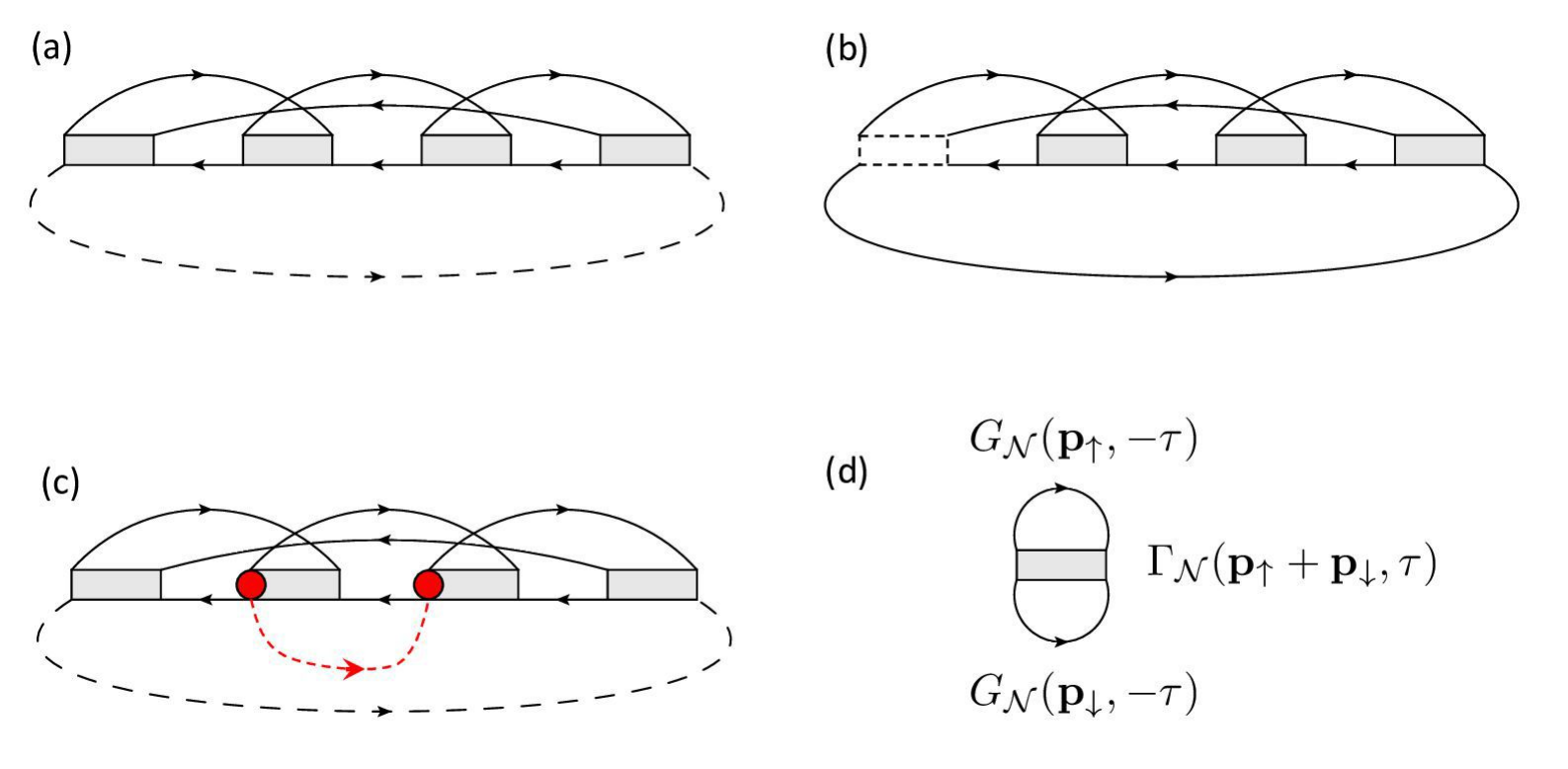}
\caption{Examples of diagrammatic topologies from the different sectors of the Monte Carlo configuration space:
(a) $\Sigma$-sector,
(b): $\Pi$-sector,
(c): Worm sector,
(d): Normalization sector.
The dashed black line is the measuring line, which has the structure of a one-body propagator in the $\Sigma$-sector, and of a pair propagator in the $\Pi$-sector. In the Worm sector, the worm ends are represented by red dots connected with an extra unphysical thread (dashed red line).
}
\label{fig:sectors}
\end{figure}

To be more specific, the allowed diagram topologies, $\Tr$, belong
to one of the following sectors (see Fig.~\ref{fig:sectors} for examples):
\bi
\item
$\Sigma$-sectors ($\Sr_{\Sigma_\sigma^{{\bl (N)}}}$):
A self-energy diagram of order $N$ contains
$N$  pair propagators $\Gamma$,
 $N-1$ single-particle propagators $G_\sigma$, 
 and $N$ single-particle propagators $G_{-\sigma}$.
The open ends of the diagram are formally closed with some extra unphysical line which
has the structure of a single-particle propagator of spin $\sigma$.
We refer to this line as the measuring line.
\item
$\Pi$-sector{\bl s} ($\Sr_{\Pi^{{\bl (N)}}}$):
A pair self-energy diagram of order $N$ contains $2N$ single-particle propagators $G$, $N-1$  pair propagators $\Gamma$, and one measuring line that has the structure of a pair propagator.
\item Worm sector:
In addition to the above physical diagrams, we also consider unphysical diagrams containing two vertices where the momentum conservation is not fulfilled. We will refer to these vertices as Worms,  named Ira ($I$) and Masha ($M$).  The momentum conservation at $I$ and $M$ is restored if we consider that
a momentum $\ddelta$ is flowing from Ira to Masha along some extra unphysical thread.
In this sector, $\Tr$ includes the location of $I$ and $M$, while the momentum $\ddelta$ is included in the internal variables $X$.
\item Normalization-sector ($\Sr_\Nr$):
The topology and variables of the normalization diagram are the ones of a fully closed $N=1$ diagram.
The lines in this diagram are certain ``designed" simple functions rather than
$G$ and $\Gamma$ propagators.
\ei

\subsection{Probability density} \label{subsec:weight}

In order to precisely define the probability density $w(\Cr) d\Cr$ on the above configuration space, we first specify what we mean by $d\Cr$.
For any function $f(\Cr)$, we set
\be
\int d\Cr f(\Cr) \equiv \sum_\Tr \int dX dY f(\Tr,X,Y) \; ,
\ee
where $dY=d\mathbf{p} ~d\tau_1 d\tau_2$ and $dX$ depends on the topology $\Tr$:
it is given by Eq.~(\ref{eq:dX}) if $\Tr$ has no Worms, and by the same expression with an additional factor $d\boldsymbol{\delta}$ if $\Tr$ has a pair of Worms.

Alternatively, one can discretize the configuration space.
We emphasize that this introduces {\it arbitrarily small} discretization steps, which is really fundamentally equivalent to working with continuous variables.
 In this case,
all momentum coordinates and imaginary time are integer multiples of some arbitrarily small $\delta p$
and  $\delta\tau$.
We can write, for topologies without Worms,
\bea
\int dX\,dY f(\Tr,X,Y)  \equiv  \sum_{\tau_1,  \ldots, \tau_{2N}} \delta\tau^{2N} \nonumber \\
  \sum_{(\pp_1,\ldots,\pp_{3N})} \delta p^{3(N+1)} f(\Tr,X,Y) \; ,
\label{eq:def_discrete}
\eea
where the sum over the momenta
$(\mathbf{p}_1,\ldots,\mathbf{p}_{3N})$
of all lines (internal and measuring) is constrained by the
momentum conservation at each vertex.
For topologies with Worms,
\bea
\int dX\,dY f(\Tr,X,Y) & \equiv & \sum_{\tau_1,  \ldots, \tau_{2N}} \delta\tau^{2N} \sum_{(\pp_1,\ldots,\pp_{3N})} \delta p^{3(N+1)} \nonumber \\
& &  \sum_{\ddelta} \delta p^3 f(\Tr,X,Y)\; .
\label{eq:def_discrete_W}
\eea
We then have
\be
\int d\Cr f(\Cr) \equiv \sum_{\Cr} \delta\Cr f(\Cr) \; ,
\ee
where $\delta\Cr$ is the ``volume" of one discrete ``cell" of the configuration space around the considered
point $\Cr$.
More precisely,
if $\Cr$ is an order-$N$ diagram, 
$\delta\Cr$ is given by $\delta\tau^{2N}\delta p^{3(N+1)}$,
multiplied by
the additional factor $\delta p^3$
if the Worms are  present.
A nice feature of the formulation~(\ref{eq:def_discrete},\ref{eq:def_discrete_W}) is that there is no need to introduce the loop momenta;
instead, all momenta are treated on equal footing, which is also the case in the diagrammatic Monte Carlo code.

We define the weighting function $w(\Cr)$ in the following way.
For a physical configuration, we take
\be
w(\Cr)=|\Dr(\Cr)|\,R(\Cr)\; ,
\label{eq:reweight}
\ee
where $\Dr(\Cr)$ is given by the Feynman rules (see subsection~\ref{subsec:feyn_rules}), and $R(\Cr)$ is
an arbitrary non-negative reweighing function.
We take
\be
R(\Cr)=W_{\rm meas}^{Q}(p)\,O_N \; ,
\ee
where
$Q$ is equal to $\Sigma$ or $\Pi$ depending on the sector,
$W_{\rm meas}^{Q}(p)$ is the weight of the measuring line,
and $O_N$ is an order-dependent reweighting factor.
We choose
$W_{\rm{meas}}^\Sigma(p,\tau)$
to be $\propto 1/p^2$ for intermediate momenta (to compensate for the Jacobian),
constant for small momenta (to avoid having
rare configurations with a large weight),
and $\propto 1/p^4$ for large momenta.
This is just one of the many possible choices, subject to the condition that
sampling of diagrams with large $p$ has to be suppressed in order to have a
normalizable distribution (i.e. the total weight $\Zr$ has to be finite).
For the $\Pi$-sector, the simplest option
is $W_{\rm meas}^\Pi=\phi_\Pi W_{\rm meas}^\Sigma$ where $\phi_\Pi$ is
an optimization factor controlling the relative weights of the  $\Sigma$ and $\Pi$ sectors.

The weight of unphysical configurations (belonging to the Worm sector or to the normalization sector)
is defined as follows. Formally, the weight of configurations containing Worms is arbitrary, since they do not contribute to the self-energy. These diagrams are auxiliary and are only employed for obtaining an efficient
updating scheme. In order to have a good acceptance ratio when moving between
the physical and Worm sectors we choose the weights according to the Feynman rules for all propagator lines,
with the extra rule that the unphysical thread contributes to $w(\Cr)$ a factor $C(\boldsymbol{\delta})$, i.e.
\be
w(\Cr)=|\Dr(\Cr)|\,R(\Cr)\,C(\boldsymbol{\delta}).
\label{eq:wcworm}
\ee
The $C(\boldsymbol{\delta})$ function should be chosen to decay fast enough
at large $\delta$ to ensure that $\Zr$ is finite and includes a constant prefactor
to optimize the relative statistics of sampled diagrams with and without the Worms.

In the normalization sector, $w(\Cr)$ is a simple expression such that one can easily calculate
analytically the total weight of the normalization sector $\Sr_\Nr$,
\be
\Zr_\Nr = \int_{\Sr_\Nr} d\Cr \, w(\Cr).
\ee
We take {\bl $w(\Cr)=G_\Nr(p_\up,-\tau)\,G_\Nr(p_\down,-\tau)\,\Gamma_\Nr(p_\up+p_\down,\tau)$} with
	$G_\Nr(p,\tau)  =  {\rm exp}({-\frac{p^2}{2 \sigma_{\Nr}^2}})$
and	$\Gamma_\Nr(p,\tau)  =   \phi_{\Nr}$.
The parameters $\sigma_{\Nr}$ and $\phi_{\Nr}$ can be freely chosen and optimized.

\subsection{Measuring}\label{subsec:measuring}

We recall that we determine the function $\Qr=\Sigma_\sigma^{{\bl(N)}}$ or $\Pi^{{\bl(N)}}$ by computing its
overlaps with a set of functions $g$.
We now describe our specific choices of functions $g(p,\tau)$.
We divide the space of all $(p,\tau=\tau_1-\tau_2)$ into bins $\Br
=\Br_p \times \Br_\tau
\subset [0,p_{\rm max}] \times [0,\beta]$.
In practice, $\tau=\tau_1-\tau_2$ lies in the interval $[-\beta,\beta]$,  but thanks to the \mbox{$\beta$-(anti-)periodicity} of $\Pi$ ($\Sigma$) we only need to consider $\tau\in[0,\beta]$.
In each bin $\Br$ we define the ortho-normal sets of basis functions
$u_k(p)$ and $v_l(\tau)$ satisfying
\bea
\int_{\Br_p} d\pp\, {\rm w}(p) u_k(p) u_{k'}(p) &=& \delta_{k,k'}
\label{eq:ps_p}
\\
\int_{\Br_\tau} d\tau\, v_l(\tau) v_{l'}(\tau) &=& \delta_{l,l'}
\eea
where ${\rm w}(p)>0$.
Then the to-be-determined function $\Qr$ can be expanded in the bin $\Br$ as
\be
\Qr(p,\tau) = \sum_{k,l} \Qr_{k,l}\,u_k(p) v_l(\tau).
\ee
Setting $g(Y) = 1_{(p,\tau)\in\Br}\,{\rm w}(p)\,u_k(p)\,v_l(\tau)$,
we obtain the expansion coefficients, $\Qr_{k,l} = \int dY \Qr(Y) g(Y)$, by Monte Carlo
as explained in Subsec.~\ref{subsec:general_idea}.

We take the ${\rm w}(p)$ function in the inner product to be ${\rm w}(p)=1/(4\pi p^2)$ (except for the lowest bin, see below) so that the $u_k$'s and $v_l$'s can be chosen in the form of Legendre polynomials up to the order $2$.
The procedure becomes exact only in the limit of vanishing bin-size, but one can afford relatively large
bins compared to the case when the function $\Qr$ is approximated by a constant in each bin
(which would correspond to restricting to the polynomial of order $0$).
In the lowest momentum-bin, we chose ${\rm w}(p)=1/(4\pi)$.
The reason for this choice is to avoid having a factor $1/p^2$ in the right-hand side of Eq.~(\ref{eq:estim_A_C}), which would lead to huge contributions from rare configurations with small $p$.
The corresponding basis set of two functions is built from a constant and $p^2$.

{\bl
This choice ensures 
that for each considered $g$, not only the mean value
$\overline{A_{\Qr,g}(\Cr)} = \lim_{n\to\infty}\frac{1}{n}\sum_{i=1}^n A_{\Qr,g}(\Cr_i)$ is finite,
but the corresponding variance
$\overline{[A_{\Qr,g}(\Cr)]^2} - [\overline{A_{\Qr,g}(\Cr)}]^2$
is also finite,
where $\overline{[A_{\Qr,g}(\Cr)]^2} = \lim_{n\to\infty}\frac{1}{n}\sum_{i=1}^n[A_{\Qr,g}(\Cr_i)]^2$.
This follows from the fact that
for each bin $\Br$ (including the special case of the lowest momentum-bin),
$[A_{\Qr,g}(\Cr)]^2$ is bounded, because
$1/W_{\rm meas}^Q(p)$ and $g(Y)$ are bounded.
}

\subsection{Updates} \label{subsec:updates}

Our updating scheme shares a number of features with the one introduced for the Hubbard model in Ref.~\onlinecite{VanHoucke1}.
To sample the space of configurations with a variable number of continuous variables,
we use a Metropolis algorithm, with pairs of complementary updates.\cite{ProkSvistTupitWorm}
In addition to the complementary pairs, a number of self-complementary updates are used. While not changing the number of
continuous variables, self-complementary  updates allow us to efficiently sample  diagram topologies.
In this Section we present details of
our specific implementation of all Monte Carlo moves including expressions for their acceptance ratios.
The updates presented in \ref{par:Create}, \ref{par:Move} and \ref{par:Shift} suffice to perform the integration over internal momenta and times while keeping the order and topology of the diagram fixed.
The updates of \ref{par:recon} and \ref{par:swapmeas} change the topology of the diagram without changing the order. The updates presented in \ref{par:addremove} and \ref{par:addremoveloop} allow one to change the diagram order.
Finally the update of \ref{par:swapnorm} allows to enter and leave the normalization sector.

\subsubsection{{Create--Delete}} \label{par:Create}
\begin{figure}
\includegraphics[width=\columnwidth]{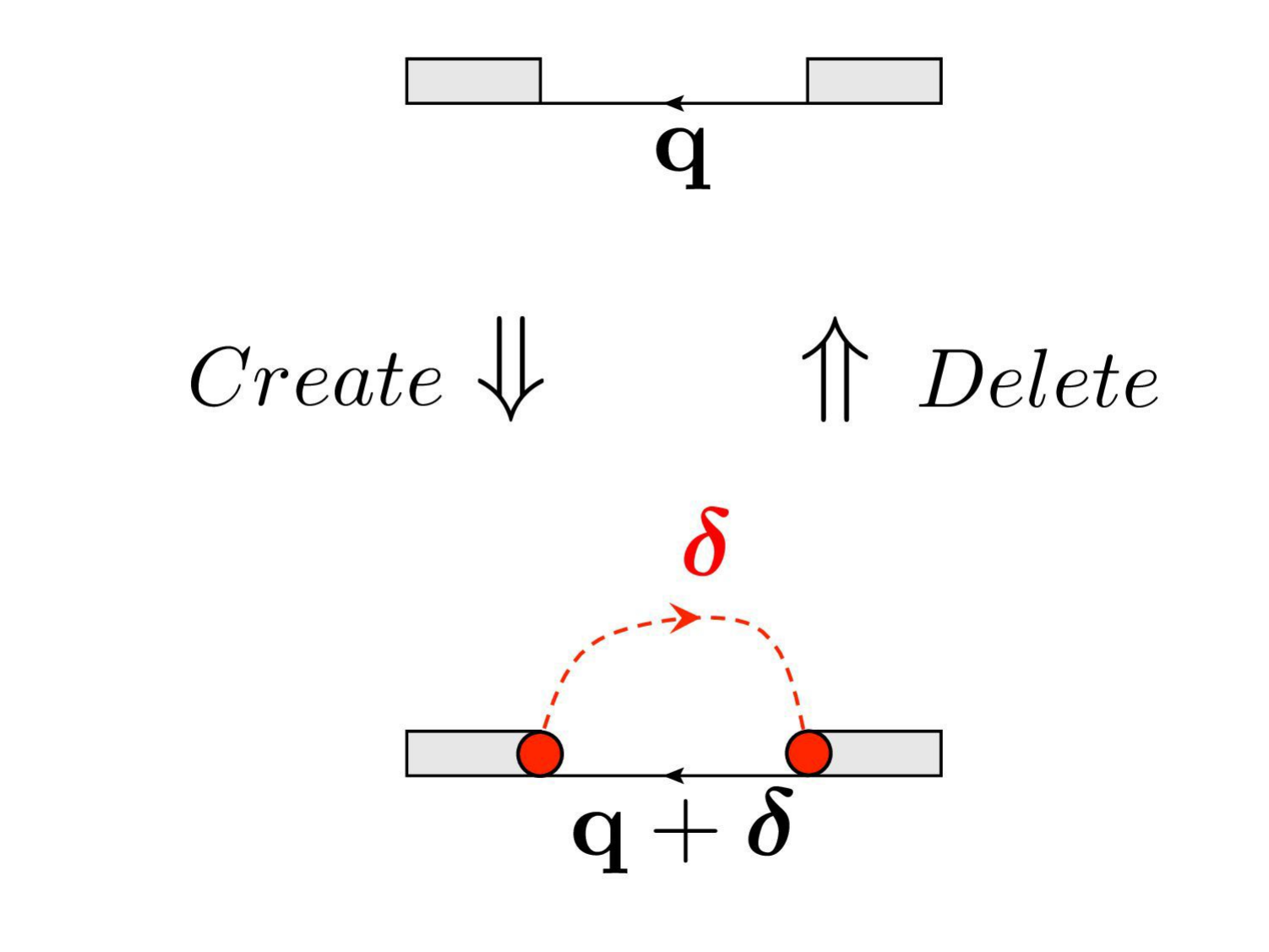}
\caption{(color online) Graphical representation of the complementary pair of updates \emph{Create-Delete}. Only relevant (i.e. updated) parts of the Feynman diagrams are drawn.
An unphysical thread (red dashed line) running from Ira to Masha and carrying momentum
$\boldsymbol{\delta}$ is added to the graph in {\it Create}, and is removed in the inverse update.
Momentum conservation is maintained when taking into account both the physical propagators
and the unphysical thread.
}
\label{fig:create}
\end{figure}

In the complementary pair of updates \emph{Create--Delete} (see Fig.~\ref{fig:create}), a pair of Worms  is created or deleted in the current diagram.
These updates are called with constant probabilities
$p_{\rm crt}$ and
$p_{\rm dlt}$, respectively.  \emph{Delete} (resp. \emph{Create}) can only be called when the Worms are present (resp. absent).
In \emph{Create}, a line is first chosen at random (i.e. with probability  $1/(3N)$ with $N$ the order of the diagram).
The chosen line can be a $G_{\sigma}$, $\Gamma$ or measuring line.
Next, we choose with equal probability either
Ira or Masha to be located at the origin of the chosen line. An unphysical thread running from Ira to Masha is introduced and carries a momentum $\boldsymbol{\delta}$,
chosen with probability density
$W(\boldsymbol{\delta})$. 
Note that, to optimize the acceptance ratio, $W$ is in principle allowed to depend on the imaginary-time
difference between the ends of the chosen line (or any other configuration parameter).

When the Worms are deleted, we first check whether there is at least one line connecting Ira and Masha.
If so, the Worms can be deleted with the inverse \emph{Create} acceptance ratio.
In case there is more than one line connecting Ira and Masha, we choose one of these lines with equal probability
$1/N_{\rm links}$ where $N_{\rm links}$ is the number of connections.
When a $G_{\sigma}$-line is chosen at the beginning of the update, the total acceptance ratio becomes
\be
q_{\rm create}\,   =  \,  \frac{p_{\rm dlt}}{p_{\rm crt}} \,
   \frac{6 N}{N_{\rm links} W(\boldsymbol{\delta}) }
 \,
\frac{ |G_{\sigma}(\mathbf{p}\pm\boldsymbol{\delta},\tau)| C(\boldsymbol{\delta})}{|G_{\sigma}(\mathbf{p},\tau)|}
 \; .
\label{eq:create}
\ee
When the chosen line is of the $\Gamma$- or measuring-type, the acceptance ratios are constructed similarly.
Recall that $C(\boldsymbol{\delta})$ is an extra factor  assigned to the diagram
with Worms [see Eq.~(\ref{eq:wcworm})]. The new momentum of the chosen line,
$\mathbf{p}+\boldsymbol{\delta}$ or $\mathbf{p}-\boldsymbol{\delta}$,
depends on whether Ira is created at the end or the origin of the line.
Finally, we are left with the choice for the probability density $W(\boldsymbol{\delta})$. 
We simply take 
$W(\boldsymbol{\delta}) \propto C(\boldsymbol{\delta})$.

\subsubsection{{Move}} \label{par:Move}

\begin{figure}
\includegraphics[width=\columnwidth]{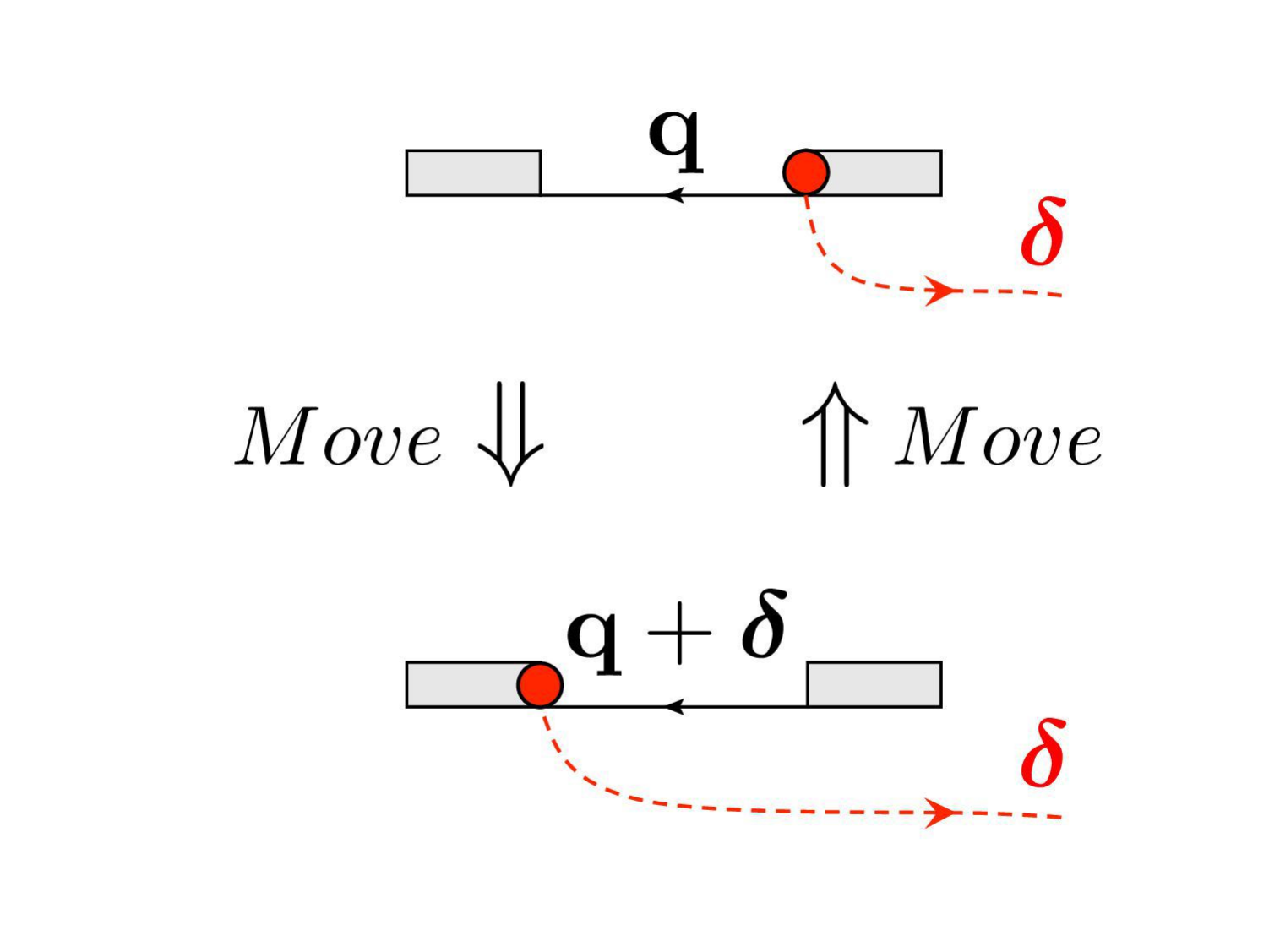}
\caption{Graphical representation of the self-complementary update \emph{Move}.  In this particular case, Ira is moved from one vertex to another along a $G_{\sigma}$-propagator line. To preserve momentum conservation, the momentum of this line is changed.
}
\label{fig:move}
\end{figure}

In \emph{Move} (see Fig.~{\ref{fig:move}}), one of the Worms is moved from
one three-point vertex to another along a single line.
This line is chosen at random (i.e., with probability $1/3$) and one has to
ensure that Ira and Masha will not be placed on the same vertex
(note that for this reason, \emph{Move} is impossible for $N=1$).
\emph{Move} is called with constant probability
$p_{\rm{move}}$, whenever Worms are present.
If a Worm happens to move along a $G_{\sigma}$-line, the acceptance ratio is
\begin{equation}
q_{\rm move} = \frac{ |G_{\sigma} (\mathbf{p} \pm \boldsymbol{\delta}, \tau)|}{ |G_{\sigma} (\mathbf{p}, \tau)|} \; .
\end{equation}
The sign depends on the direction in which Ira or Masha is moved,
and is chosen such that momentum conservation is preserved.
As a result of {\it Move} updates, Ira and Masha perform a random work
within the Feynman graph, while constantly updating line momenta.

\subsubsection{{Shift in time}} \label{par:Shift}

\begin{figure}
\includegraphics[width=\columnwidth]{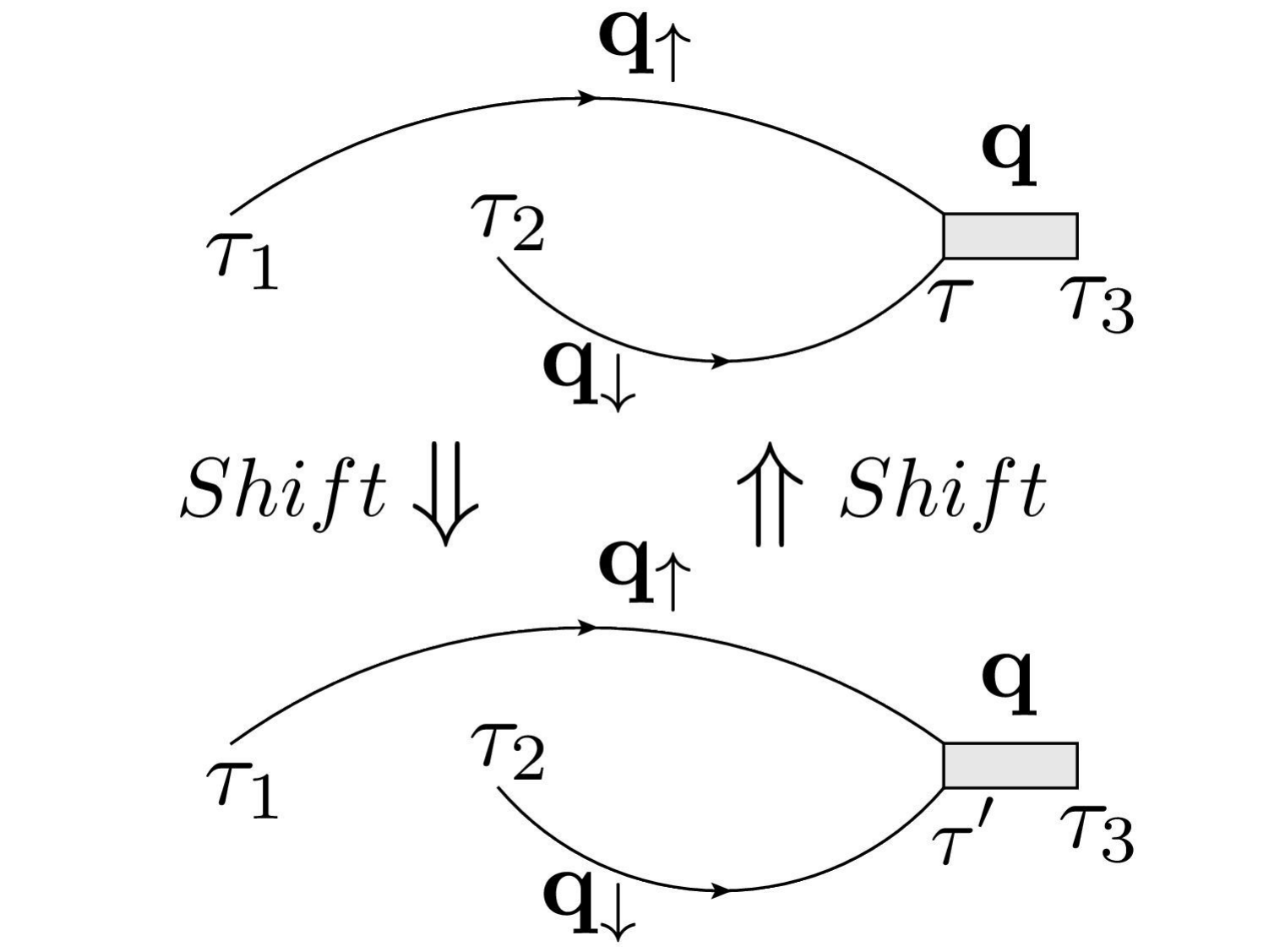}
\caption{Graphical representation of the self-complementary update \emph{Shift}.
The time $\tau$ of a three-point vertex is shifted to $\tau'$.
}
\label{fig:shift}
\end{figure}

This update can be called in every sector with constant probability
$p_{\rm shift}$.
It shifts the time variable of the randomly selected three-point vertex from $\tau $ to $\tau'$.
The new variable $\tau'$ is drawn from a distribution $W(\tau')$ 
{\bl on the $(0,\beta)$ interval}.
The acceptance ratio is
(here it is given for the case when all three lines attached to the shifted vertex
are physical propagators)
\bml
q_{\rm shift}    =    \frac{W(\tau)  }{W(\tau')    } \\
\cdot \,  \frac{  |G_{\uparrow}(\mathbf{q}_{\uparrow},\tau'-\tau_{1})   G_{\downarrow}(\mathbf{q}_{\downarrow},\tau'-\tau_{2})  \Gamma(\mathbf{q},\tau_3 - \tau')|}{   |G_{\uparrow}(\mathbf{q}_{\uparrow},\tau-\tau_{1})   G_{\downarrow}(\mathbf{q}_{\downarrow},\tau-\tau_{2})  \Gamma(\mathbf{q},\tau_3 - \tau)| }
\; ,
\end{multline}
where we have assumed that the propagators $G_{\sigma}$ are incoming (see Fig.~\ref{fig:shift}).
{\bl
We choose a seeding function $W$
taking into account the short-time behavior $\Gamma(k,\tau) \propto 1/\sqrt{\tau}$
[see Eqs.~(\ref{eq:Gamma_large_k},\ref{eq:def_Gamma_c}) below]:
$W(\tau')= 1/(2\sqrt{\beta\Delta\tau})$
where $\Delta\tau\equiv\tau_3-\tau'$ for $\tau_3>\tau'$,
and $\Delta\tau\equiv\tau_3-\tau'+\beta$ for $\tau_3<\tau'$.
}

\subsubsection{{Reconnect}}\label{par:recon}

This self-complementary update changes the topology of the diagram without changing the order.
It is called with constant probability
$p_{\rm rec}$
whenever  Worms  are present and $N>1$.
The basic idea is that two spin-$\sigma$ single-particle propagators that both leave from (or both arrive at) the Worm-vertices are reconnected, i.e. their end-points are exchanged. 
 This
does not cause a problem with momentum conservation: Only the unphysical momentum $\boldsymbol{\delta}$ running from Ira to Masha changes.

\begin{figure}
\includegraphics[width=\columnwidth]{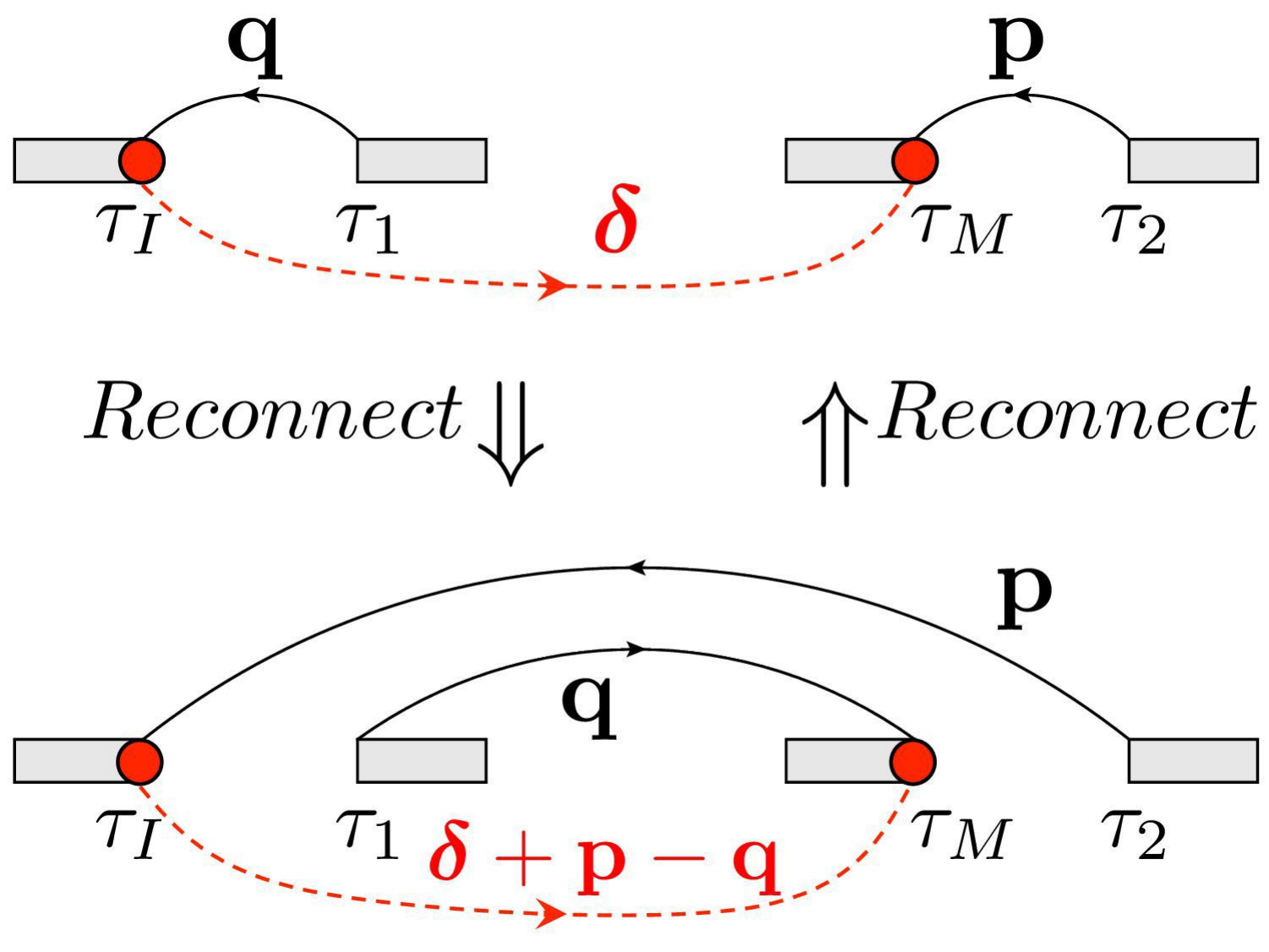}
\caption{Graphical representation of the self-complementary update \emph{Reconnect}.  In this particular example, the two propagators incoming to the Ira and Masha-vertices are interchanged.
The momentum carried by the unphysical thread is changed from $\boldsymbol{\delta}$ to $\boldsymbol{\delta}+\mathbf{p}-\mathbf{q}$.
}
\label{fig:reconnect}
\end{figure}

\emph{Reconnect} is constructed as follows (see Fig.~\ref{fig:reconnect}).
First, we choose with equal probability whether to reconnect the spin-up lines or the spin-down lines.
These propagators should be both  arriving at or both leaving from the Worms, otherwise the update is rejected.
In case they both arrive at the Worms, the acceptance ratio is
\bea
q_{\rm{reconnect}} & = &  \frac{ | G_{\sigma}(\mathbf{q}, \tau_M-\tau_1) G_{\sigma}(\mathbf{p}, \tau_I-\tau_2) |    }{| G_{\sigma}(\mathbf{q}, \tau_I-\tau_1) G_{\sigma}(\mathbf{p}, \tau_M-\tau_2) |  } \;  \nonumber \\
& & \cdot \frac{C(\boldsymbol{\delta}+\mathbf{p}-\mathbf{q}) }{C(\boldsymbol{\delta})} .
\eea

\subsubsection{{Swap measuring line}}\label{par:swapmeas}

\begin{figure}
\includegraphics[width=0.8\columnwidth]{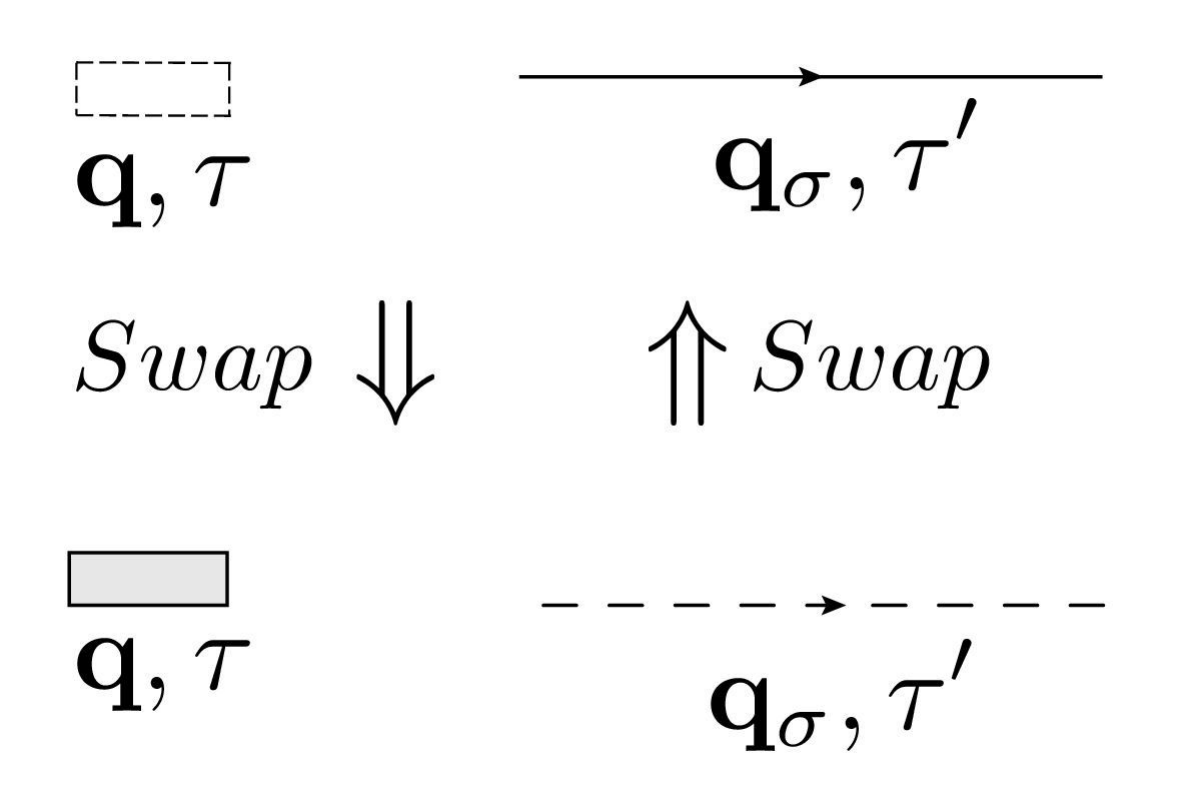}
\caption{Graphical representation of the \emph{Swap measuring line} update.
}
\label{fig:swap}
\end{figure}

This update
converts the measuring line into a real propagator,
while some other line becomes the new measuring line
 (see Fig.~\ref{fig:swap}).
 Although very simple, this update
  changes the diagram topology and the
values of internal and external variables.  The update is only called in the $\Sigma$ and $\Pi$ sectors,
since it is not useful in the Worm or normalization-sector. The update starts with choosing one of the lines at random (it should not be the measuring line). This line is proposed to become the new measuring line.
The acceptance ratio is given by
\begin{equation}
q_{\rm{swap}}  =  \frac{|\Gamma(\mathbf{q}, \tau)| }{W_{\rm{meas}}^\Pi(\mathbf{q}, \tau) }  \cdot \frac{W_{\rm{meas}}^\Sigma(\mathbf{q}_{\sigma}, \tau')}{|G_{\sigma}(\mathbf{q}_{\sigma}, \tau')|} \; ,
\end{equation}
for the particular case which converts $\Pi$-sector to $\Sigma$-sector.
For other cases, acceptance ratios are constructed similarly.

\subsubsection{{Add--Remove}} \label{par:addremove}

\begin{figure}
\includegraphics[width=0.8\columnwidth]{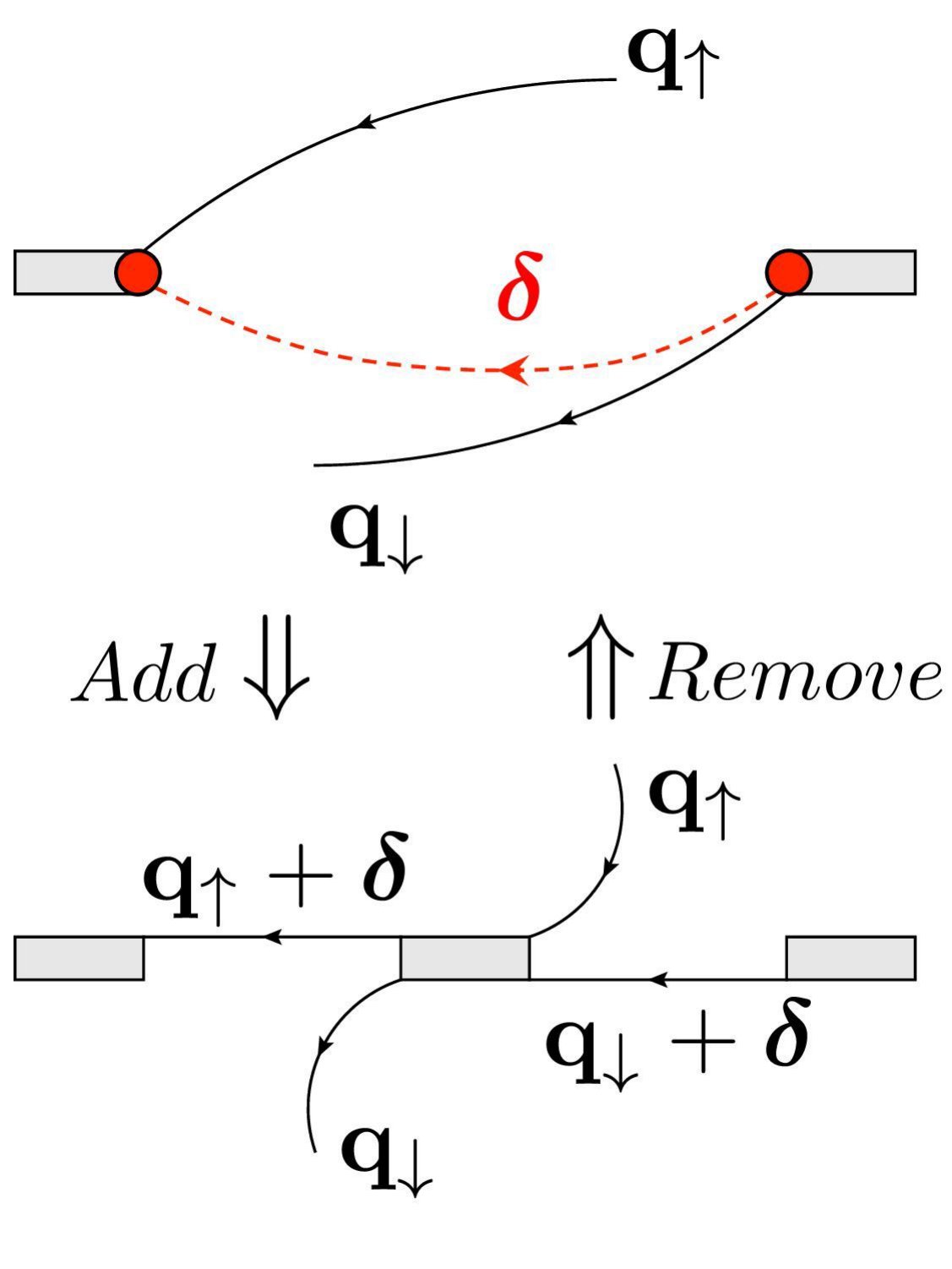}
\caption{Graphical representation of the complementary pair of moves \emph{Add-Remove}.  In this particular example, the $G_{\sigma}$ propagator in Eq.~(\ref{eq:qadd}) corresponds to the spin-$\downarrow$ line carrying momentum $\mathbf{q}_{\downarrow}$, and $G_{-\sigma}$ is the spin-$\uparrow$ line with momentum $\mathbf{q}_{\uparrow}$.
}
\label{fig:add2}
\end{figure}

To add a pair-propagator line, the Worms should be present, and we should not be dealing with the normalization diagram.
In this case, the update \emph{Add} is called with constant probability
$p_{\rm add}$.
First, we choose the spin-up or spin-down line attached to the Ira-vertex.  Let this line correspond to $G_{\sigma}$. Next, we consider
the opposite spin propagator attached to the Masha-vertex, $G_{-\sigma}$.
These two propagators will be cut, and a new pair propagator will be inserted;
see Fig.~\ref{fig:add2} for an illustration.
The final diagram does not contain the Worms, which leaves us no freedom in choosing
the momenta in the final diagram.
We propose initial and final times $\tau_o$ and $\tau_d$ for the new pair-propagator line,
from a probability density $W(\tau_o, \tau_d)$, which in the current implementation is simply the uniform distribution.

For \emph{Remove}, we need $N>1$ and the Worms should be absent.
The update is called with probability $p_{\rm rm}$.
The pair-propagator line to be removed is chosen at random.
If the topology of the diagram is such that the chosen $\Gamma$-line has the same
$G$-propagator attached to its both ends, the update is immediately rejected
since such a $G$-loop cannot be created through \emph{Add}.
An update trying to remove a measuring $\Gamma$-line
is also forbidden.
Next,  choose one of the four lines attached to the pair-propagator line at random.
This will be the future $G_{\sigma}$-line and the vertex it is connected to will
become Ira. One of the remaining $G_{-\sigma}$ is also selected at random and
the vertex it is connected to will become Masha. If the same vertex is chosen
for Ira and Masha, the move is rejected.

The acceptance ratio for \emph{Add} is
\bml
	q_{\rm add}\,  = \, \frac{p_{\rm rm}}{p_{\rm add}} \, \frac{   O_{N+1}   \Gamma(\tau_d - \tau_o) }
	{32\pi^3  (N+1) W(\tau_o, \tau_d)   ~C(\boldsymbol{\delta})   ~O_N}
	\\
	\frac{ G_{\sigma}(\tau_{o}- \tau_{1o}) G_{-\sigma}(\tau_{o}- \tau_{2o})  G_{\sigma}(\tau_{1d}- \tau_{d}) G_{-\sigma}(\tau_{2d}- \tau_{d})}{G_{\sigma}(\tau_{1d}- \tau_{1o}) G_{-\sigma}(\tau_{2d}- \tau_{2o})}
	\;.
\label{eq:qadd}
\end{multline}
The momenta are omitted here for simplicity. There are several possibilities depending on the particular choice of $G_{\sigma}$ and $G_{-\sigma}$ and the positions of Ira and Masha. In all cases, however,
the new momenta are completely determined by the conservation laws. Fig.~\ref{fig:add2} shows a particular example.
If in {\it Add} the chosen propagator $G_{\sigma}$ (or $G_{-\sigma}$) happens to be the measuring line,
then a new measuring line will be chosen with equal probability among the two
spin-$\sigma$ propagators connected to $\Gamma$ in the final diagram.
The reverse is done in \emph{Remove}.

\subsubsection{{Add--Remove loop}} \label{par:addremoveloop}

\begin{figure}
\includegraphics[width=0.7\columnwidth]{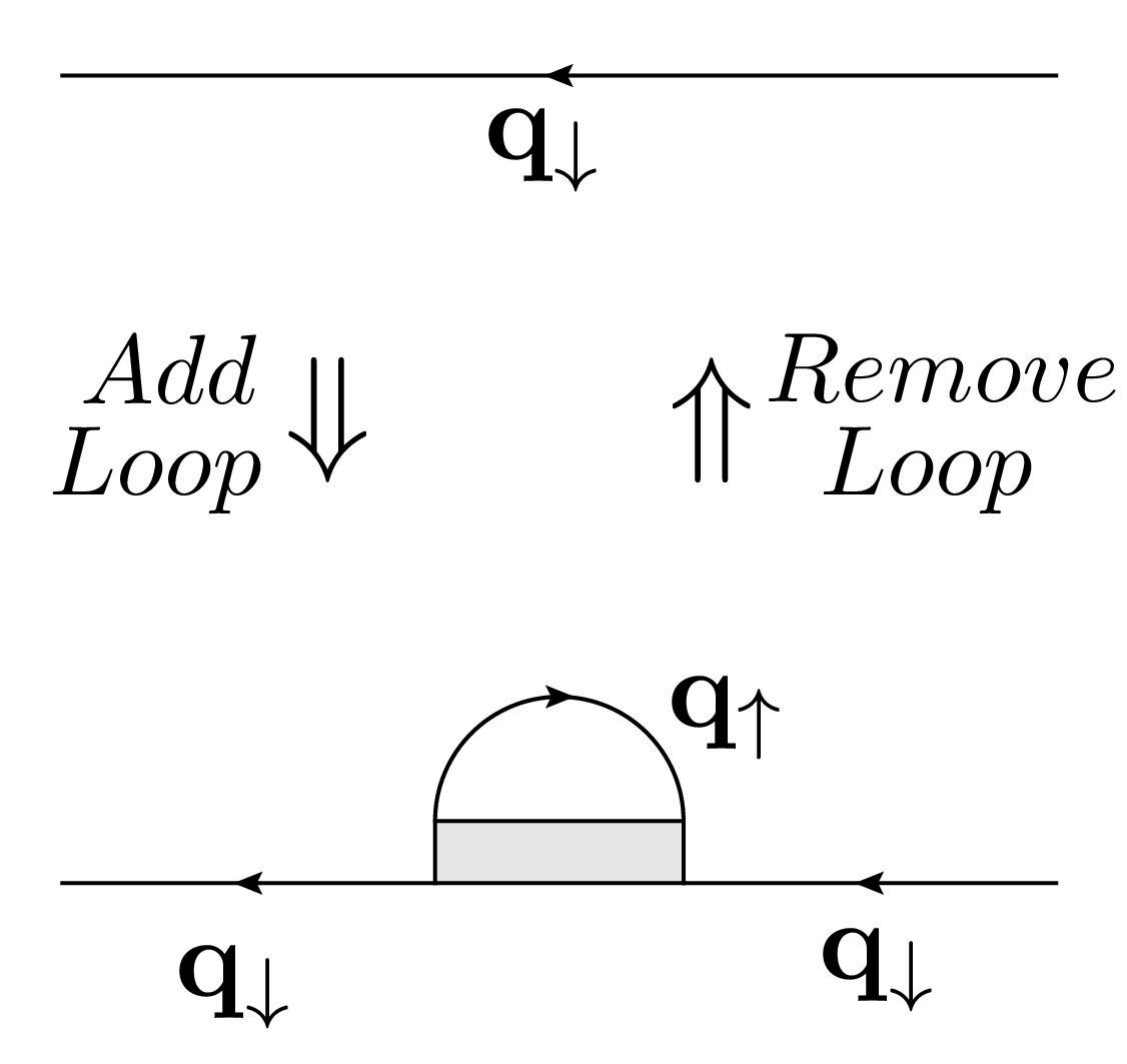}
\caption{Graphical representation of the updates \emph{Add--Remove loop}.
}
\label{fig:addloop}
\end{figure}

These updates are called in the $\Sigma$- and $\Pi$-sectors only.
\emph{Add loop} (resp. \emph{Remove loop}) is called with the probability $p_{\rm al}$ (resp. $p_{\rm rl}$).
In \emph{Add loop} a $G_{\sigma}$ propagator is chosen at random, and converted into the sequence $G_{\sigma}\Sigma_{\sigma}^{(1)} G_{\sigma}$ where $\Sigma_{\sigma}^{(1)}$ is the first-order
self-energy diagram ($\Gamma$ closed with $G_{-\sigma}$), see Fig.~\ref{fig:addloop} where we illustrate
the setup for $\sigma =\ \downarrow$ .
The initial and final times $\tau_o$ and $\tau_d$ for the new pair-propagator line are chosen from the
probability density $W(\tau_o,\tau_d)$,
and the momentum $\mathbf{q}_{\uparrow}$ for $G_{\uparrow}$ is chosen from another distribution $W(\mathbf{q}_{\uparrow}|\tau_o-\tau_d)$.
In \emph{Remove loop} a pair-propagator line is first chosen at random.
If this propagator has the same $G$-line attached to its ends, then it can
possibly be removed by the update.  If either the $\Gamma$ or $G_{\uparrow}$-line
is the measuring line, the update is rejected.
The acceptance ratio for \emph{Add loop} is given by
\begin{eqnarray}
q_{{\rm add\, loop}}  & = &    \frac{p_{\rm rl}}{p_{\rm al}} \,  \frac{(2N-1) \, O_{N+1}}{(2\pi)^3 (N+1)  W(\tau_o,\tau_d) W(\mathbf{q}_{\uparrow}|\tau_o-\tau_d) O_N} \nonumber \\
& & \cdot ~\frac{|G_{\downarrow}(\mathbf{q}_{\downarrow}, \tau_2-\tau_d)G_{\downarrow}(\mathbf{q}_{\downarrow}, \tau_o-\tau_1) |}{|G_{\downarrow}(\mathbf{q}_{\downarrow}, \tau_2-\tau_1)|} \nonumber \\
& &  \cdot ~ |\Gamma(\mathbf{q}_{\downarrow}+\mathbf{q}_{\uparrow}, \tau_d-\tau_o) G_{\uparrow}(\mathbf{q}_{\uparrow}, \tau_o-\tau_d)| \; ,
\end{eqnarray}
with $N$ the order of the diagram in \emph{Add loop}. For $W(\mathbf{q}_{\uparrow}|\tau_o-\tau_d)$
we take a Gaussian distribution with variance $1/\tau$ where $\tau=\tau_o-\tau_d$ for $\tau_o>\tau_d$ and $\tau=\tau_o-\tau_d+\beta$ for $\tau_o<\tau_d$.
This corresponds to the behavior of the vacuum propagator $G_v$.

\subsubsection{{Swap to the normalization diagram}} \label{par:swapnorm}

For normalization purposes, we introduce an unphysical diagram, for which all integrals can be evaluated analytically (see subsections \ref{subsec:config} and \ref{subsec:weight}).
If the current diagram is the one-body self-energy diagram of order one, the \emph{Norm} update proposes to swap to the normalization diagram.
The acceptance ratio is
\begin{equation}
	q_{Norm} = \frac{G_\Nr(\mathbf{p}_\up,-\tau)G_\Nr(\mathbf{p}_\down,-\tau)\Gamma_\Nr(\mathbf{p}_\up+\mathbf{p}_\down,\tau)
	 }{|W_{\rm meas}^{\Sigma}(\mathbf{p}_\sigma,-\tau)G_{-\sigma}(\mathbf{p}_{-\sigma},-\tau)\Gamma(\mathbf{p}_\up+\mathbf{p}_\down,\tau)|}.
	\label{eq:swapnorm}
\end{equation}

When the current diagram is the normalization diagram, \emph{Norm} proposes to swap back to the physical self-energy diagram with the probability given by the inverse of Eq.~(\ref{eq:swapnorm}).

\subsection{Reducibility and ergodicity}\label{subsec:ergodic}

The goal of our Monte Carlo setup is to sample the space of one-body and two-body self-energy skeleton diagrams in an ergodic way. These diagrams are connected, irreducible with respect to cutting a single $G$-propagator or $\Gamma$-propagator, and irreducible with respect to cutting
any two $G_{\sigma}$-propagators
 or any two $\Gamma$-propagators.
The set of updates presented in Section~\ref{subsec:updates} suffices to generate this class of diagrams.
In principle, the scheme could  be used to generate a bigger class of diagrams
(e.g., all connected diagrams), but we focus the discussion here on sampling the skeleton diagrams only.

Some of the updates of  Section~\ref{subsec:updates} can propose to go from a skeleton diagram to a non-skeleton one.
One possibility is that all such proposals are simply rejected.
This immediately creates a problem with ergodicity: Since there is no skeleton diagram at order 2
and the diagram order can only be changed by one, the simulation would never leave the first-order diagram.  Allowing some non-skeleton diagrams at orders 2 and 3 solves the problem and is sufficient for ergodicity
(obviously, non-skeleton diagrams are excluded from the measurements).
Beyond order 3, we restrict sampling to skeleton diagrams only without violating the ergodicity requirement.

Explicitly checking the topology of high-order diagrams at each update would be very time-consuming.
Instead, our connectivity and reducibility checks rely on momentum conservation.
Let us start with discussing the connectedness of the generated diagram. It is easy to see that, by construction, the only moves that can possibly generate disconnected pieces are \emph{Reconnect} and \emph{Remove}.
The latter update, however, can only create a disconnected piece if the initial diagram is not a skeleton diagram (since this diagram falls apart when cutting  two $G_{\sigma}$ lines connected to the $\Gamma$-line that is removed).
\emph{Reconnect}, on the other hand, can generate  two disconnected pieces in the Worm sector starting from a skeleton diagram. We simply reject the update when this happens, which can be straightforwardly done in the following way. When two disconnected pieces are generated by \emph{Reconnect}, the  Worms will be located on two three-point vertices which are part of these two pieces. Due to momentum conservation, $\boldsymbol{\delta} = 0$. 
In this case, we reject the update.

To test the topology of the diagram, we keep momenta of all lines in a hash table.
The key point is that a diagram has
an irreducible skeleton topology if and only if  no pair of lines 
(irrespective of the type of line: $G$, $\Gamma$, or measuring line)
can have exactly the same momentum 
(or momenta which differ by $\pm\boldsymbol{\delta}$ in a Worm sector)
with finite probability.
Indeed, such a pair of lines can only exist if
the two lines are of the same type and if the diagram falls apart when cutting these two lines.
The hash table allows to find equal momenta in just a few operations
for a sufficiently fine mesh in the hash table. Whenever a momentum of a line is changed, the hash table is updated.
In each update, we ensure that the final diagram will be of the skeleton type.
Note that many of the updates cannot, by construction, result in disconnected or non-skeleton topology, and
we only check these properties when there is a possibility that such a topology will be created.
For example, 
when adding a pair-propagator in \emph{Add}, there is only one way in which the diagram can become non-skeleton: when the final diagram
falls apart by cutting the added pair-propagator and another line. This means that the added pair-propagator will
be having the same momentum than another line.

We have checked ergodicity explicitly using a dedicated program which enumerates all topologies.
In practice, we ran these checks up to order $8$.
As a byproduct we get the number of topologies at each order, given in Table~\ref{tab:nb_diag}.~\footnote{These numbers were used and partially checked in Ref.~\onlinecite{CountingFeynDiagKugler}.}
As mentioned earlier, we allow some non-skeleton diagrams at order 2 and 3 to ensure ergodicity, namely the one-particle irreducible diagrams without ladders (their number is also given in Table~\ref{tab:nb_diag}).
For this reason we have introduced the moves  \emph{Add Loop}   and \emph{Remove Loop} that add and remove loops.
These updates should not be called 
if  the final (initial) order is bigger than 3.

\begin{table}

\begin{tabular}{|c||c|c|c|c|c|c|}
	\hline
N  &   $\Sigma_\sigma[G,\Gamma]$ &  $\Pi[G,\Gamma]$ & $\Sigma_\sigma[G^{(0)},\Gamma^{(0)}]$ & $\Pi[G^{(0)},\Gamma^{(0)}]$  \\
	\hline \hline
1   		& 1             	& 1	& 1	           & 0			 	\\
2   		& 0      		& 0	& 1   		  & 2 				\\
3   		& 1    		& 1 	& 5   		  & 6 				\\
4   		& 4   			& 4	&  25    	  & 30 			\\
5  		& 23			& 23	&    161  	  & 186 			\\
6  		& 168 		 & 168 	&  1201   	  & 1362 			\\
7        	& 1384		& 1384	& 10181	  & 11382   		\\
8  		& 12948		& 12948	&  96265	  & 106446  		\\
	\hline
\end{tabular}

\caption{Number of {\bl diagram topologies} contributing to the one-body self-energy $\Sigma_\sigma$ and two-body self-energy $\Pi$. 
In addition to the number of skeleton diagrams built with $G$ and $\Gamma$
(first and second column),
we also give for comparison the number of diagrams built with $G^{(0)}$ and $\Gamma^{(0)}$
(third and fourth column).
\label{tab:nb_diag}}
\end{table}

\subsection{Bold diagrammatic Monte Carlo iterative scheme} \label{subsec:iter}

The self-consistent nature of BDMC implies that the calculation is performed iteratively.
Starting from the propagators $G$ and $\Gamma$ (for the first iteration, they are just some initial guess), the self-energies $\Sigma$ and $\Pi$ are calculated by diagrammatic Monte Carlo. They are used next
in the Dyson equations to compute new values of the propagators, and the simulation continues with updated
propagator lines.
After a large enough number of iterations, the process converges. 

A useful trick to accelerate this convergence is to perform a weighted average over different iterations.\cite{ProkofevSvistunovBoldPRL} More precisely, the self-energy $\Sigma_j$ that we plug into the Dyson equation after iteration $j$ is a weighted average of the Monte Carlo result 
of  iteration $j$, and of $\Sigma_{j-1}$. The corresponding weighting coefficients can be optimized to obtain small statistical errors as well as fast $j$-dependent convergence.

{\bl We have used the following weighting coefficients.
The Monte Carlo result of iteration $j$ is an unnormalized histogram
 $\bar{\Sigma}_j^{(h)}$
and a norm $\bar{\Nr}_j$; 
instead of
estimating the self-energy as
$\Sigma_j=\bar{\Sigma}_j^{(h)}/\bar{\Nr}_j$,
we use 
$\Sigma_j = \Sigma_j^{(h)} / \Nr_j$
with $\Sigma_j^{(h)} = \bar{\Sigma}^{(h)}_j + (1-f_j) \Sigma_{j-1}^{(h)}$
and $\Nr_j = \bar{\Nr}_j + (1-f_j) \Nr_{j-1}$.~\footnote{The relation with the notations of Sec.~\ref{subsec:general_idea} is:
$\bar{\Nr}_j = \Nr = \sum_{i=1}^n 1_{\Cr_i\in\Sr_\Nr}$
and
$\bar{\Sigma}_j^{(h)} = Z_\Nr\ \sum_{i=1}^n A_{\Sigma_\sigma^{(N)},g}(\Cr_i)$, with $n$ the number of Monte~Carlo steps per iteration.}
Using a non-zero value of $f_j$ suppresses the contribution of older iterations,
leading to a faster convergence.~\footnote{This was demonstrated to us by E. Kozik.} In the long-time limit $j\to\infty$, the statistical error still tends to zero provided $f_j\to0$.
We used $f_j\propto1/j$.
The same procedure was applied for $\Pi$.
}

{\bl
The final error for each observable (density, contact, etc.) was estimated conservatively from its fluctuations as a function of the iteration number $j$.
More precisely, 
for a total number $J$ of iterations,
we estimated the error as the maximal deviation between
the final result after iteration $J$
and all intermediate  results after the iterations $j\in [J/2, J]$.
This automatically takes into account the combined effects of the statistical errors and the error due to the finite number of self-consistency loops.}

\subsection{Resummation} 
\label{sec:resum}

If the diagrammatic series was convergent, we would simply have
\be
Q=\lim_{N_{\rm max}\to\infty} \ \sum_{N=1}^{N_{\rm max}} Q^{(N)} \, ,
\label{eq:no_resum}
\ee
where
$Q$ stands either for the single-particle self-energy $\Sigma$ or for the pair self-energy $\Pi$,
 $Q^{(N)}$ is the total contribution of the $N$-th order diagrams,
and where it is implicit that we consider arbitrary fixed values of the external variables $(p,\tau)$.
{\bl However, the series is divergent, as shown analytically in Refs.~\onlinecite{RossiEOS,ResonLong2}.
To overcome this difficulty, we employ a divergent-series-resummation method, of the form
\be
Q=\lim_{N_{\rm max}\to\infty}\  \sum_{N=1}^{N_{\rm max}} R_N^{(N_{\rm max})}\, Q^{(N)}
\label{eq:resum}
\ee
where the $R_N^{(N_{\rm max})}$ are appropriate coefficients,
{\bl corresponding to a conformal-Borel transformation,}
 see Refs.~\onlinecite{RossiEOS,ResonLong2}.

In practice, a full BDMC calculation must be performed for each value of $N_{\rm max}$, and the result is extrapolated to $N_{\rm max}\to\infty$.
This implies that the $Q^{(N)}$ are themselves $N_{\rm max}$-dependent, and are assumed to tend to the exact $Q^{(N)}$ when the $N_{\rm max}\to\infty$ limit is taken.
}

\section{Ultraviolet physics}\label{sec:UV}

Zero-range interactions lead to a characteristic ultraviolet asymptotic behavior governed by the so-called contact.\cite{TanEnergetics,TanLargeMomentum,ChapLeggettBref,ChapBraatenBref,LeChapitreBref}
This physics is expressed in a natural way within the bold-line diagrammatic framework, 
as we explain in subsection~\ref{subsec:analytics}
(related discussions within the $T$-matrix approximation can be found in Refs.~\onlinecite{CombescotC,StrinatiRF,ZwergerRFLong,Hu_C}).
Analytical understanding of the ultra-violet behavior is readily incorporated
into our BDMC scheme, as described in subsection~\ref{subsec:incorp}.
A short description of these points was given in Ref.~\onlinecite{RossiContact}.

\subsection{Large-momentum analytics} \label{subsec:analytics}

\subsubsection{The contact}

The momentum distribution of the resonant gas has the power-law tail
\be
n_{\sigma}(k) \underset{k\to\infty}{\sim} \frac{\Cr}{k^4}.
\label{eq:C_nk}
\ee
In practice, this behavior holds for $k$ much larger than
the typical momentum $k_{\rm typ}$ of the particles in the gas.
(In the balanced unitary case, $k_{\rm typ}$ is the maximum of the Fermi momentum and the thermal momentum.)

In position space, the density-density correlation function  diverges at short distance as
\be
\langle \, \hat{n}_\up(\rr) \,
\hat{n}_\down(\vn) \,
 \rangle
\underset{r\to0}{\sim}\frac{\Cr}{(4\pi\, r)^2}.
\label{eq:C_g2}
\ee
An immediate consequence of the last equation is that
if one measures all the particle positions
in a unit volume,
the number of pairs of particles whose interparticle distance is smaller than $s$ is $\Cr s/(4\pi)$ when $s\to0$;
in this sense, $\Cr$ can be viewed as a density of short-distance pairs.\cite{TanEnergetics,BraatenC,ChapBraatenBref}

Furthermore, the contact can be directly expressed in terms
of the bold pair propagator
\be
\Cr = - \Gamma(\rr=\vn,\tau=0^-).
\label{eq:CvsGamma}
\ee
This expression is analogous to the expression Eq.~(\ref{eq:n_vs_G}) of the single-particle density $n$ in terms of the single-particle propagator $G$, which shows again that $\Cr$ controls the density of short-distance pairs.
While Eq.~(\ref{eq:CvsGamma}) was first obtained within the
$T$-matrix approximations,\cite{StrinatiRF,ZwergerRFLong,Hu_C} it is actually
an exact relation in terms of the fully dressed $\Gamma$.\cite{RossiContact}

\subsubsection{Bold propagators at large momentum} \label{sec:G_Gamma_large_k}

At large momentum, the bold propagators can in some sense be replaced by vacuum propagators.
More precisely,
when $k\to\infty$,
$G(k,\tau)$ and $\Gamma(k,\tau)$ become small for any $\tau$ in the interval $]0;\beta[$, except in the narrow region $0<\tau\lesssim 1/k^2$ where
\bea
G(k,\tau)
&\simeq&
G_v(k,\tau) \label{eq:G_large_k}
\\
\Gamma(k,\tau)
& \simeq &
 \Gamma_v(k,\tau)
\label{eq:Gamma_large_k}
\eea
with
\bea
G_v(k,\tau) & \equiv& - e^{-(k^2/2)\tau}
 \label{eq:def_G_v}
\\
 \Gamma_v(k,\tau) &\equiv& - 4 \sqrt{\frac{\pi}{\tau}} \ e^{-(k^2/4)\tau}.
 \label{eq:def_Gamma_c}
 \eea
This can be justified as follows. We first note that
$G^{(0)}(k,\tau)\simeq G_v(k,\tau)$ at large $k$,
where we extend $G_v$ to negative times by $\beta$-antiperiodicity.
To justify
(\ref{eq:G_large_k}), we write $(G-G^{(0)})(k,\tau) = (G^{(0)}\Sigma G^{(0)})(k,\tau) + \ldots
=
\int_0^\beta d\tau_1
\int_0^\beta d\tau_2
\, G^{(0)}(k,\tau-\tau_1)\Sigma(k,\tau_1-\tau_2) G^{(0)}(k,\tau_2)
+\ldots$.
 When $k\to\infty$,  $G^{(0)}(k,\Delta\tau)\simeq G_v(k,\Delta\tau)$ becomes a narrow function of $\Delta\tau$, 
 so that
 the integrals over the internal times $\tau_i$  are effectively restricted to narrow intervals of width $\sim 1/k^2$.
This implies that
$G(k,\tau)-G^{(0)}(k,\tau)$
tends to zero uniformly in $\tau$ when $k\to\infty$.

 To derive (\ref{eq:Gamma_large_k}), we first note that $\Gamma^{(0)}(k,\tau)\simeq\Gamma_v(k,\tau)$ at large $k$ and $\tau\lesssim1/k^2$,
 as shown in Appendix~\ref{ap:Gamma0_prop}.
Equation (\ref{eq:Gamma_large_k}) then follows by writing
 $(\Gamma-\Gamma^{(0)})(k,\tau) = (\Gamma^{(0)}\,\Pi\, \Gamma^{(0)})(k,\tau) + \ldots
=
\int_0^\beta d\tau_1
\int_0^\beta d\tau_2
\Gamma^{(0)}(k,\tau-\tau_1)\Pi(k,\tau_1-\tau_2) \Gamma^{(0)}(k,\tau_2)
+\ldots$.
Again, when $k\to\infty$,
the integrals over the internal times $\tau_i$  are effectively restricted to narrow intervals,
so that
$\Gamma(k,\tau)-\Gamma^{(0)}(k,\tau)$
tends to zero uniformly in $\tau$.

We  have also derived analytical expressions for $G-G^{(0)}$ at large momentum or short distance,
which naturally depend on the contact.
These expressions are given in Appendix~\ref{app:G_large_k}
and used in Appendix~\ref{app:1st_order}.

\subsubsection{Self-energy at large momentum}
When $k\to\infty$,
$\Sigma_\sigma(k,\tau)$
becomes small for any $\tau$ in the interval $]0;\beta[$, except 
\bi
\item
for $\tau\to0^+$ with $0<\tau\lesssim 1/k^2$, where
\be
\Sigma_\sigma(k,\tau)
\simeq
 \Sigma^{(+)}_\sigma(k,\tau)
 \label{eq:sig+}
 \ee
 with
 \be
  \Sigma^{(+)}_\sigma(k,\tau)
  \label{eq:def_sig+}
  \equiv
-4 \sqrt{\frac{\pi}{\tau}}\,n_{-\sigma}\,e^{-(k^2/4) \tau} 
\ee
\item
for $\tau\to\beta^-$ with $0<\beta-\tau\lesssim 1/k^2$, where
\be
\Sigma_\sigma(k,\tau) \simeq
 \Sigma^{(-)}(k,\tau)
 \label{eq:sig-}
 \ee
 with
 \be
 \Sigma^{(-)}(k,\tau) \equiv
  -\Cr\,e^{-(k^2/2)(\beta-\tau)}.
  \label{eq:def_sig-}
\ee
\ei
Furthermore, this behavior 
comes entirely from the lowest-order bold diagram $\Sigma_\sigma^{(1)}$.~\footnote{Starting from
Eq.~(3.36) of Ref.~\onlinecite{NishidaHardProbes},
 our
Eqs.~(\ref{eq:def_sig+},\ref{eq:def_sig-}) 
can be rederived
[Y. Nishida, private communication].
}

\begin{figure}
\includegraphics[width=0.7\columnwidth]{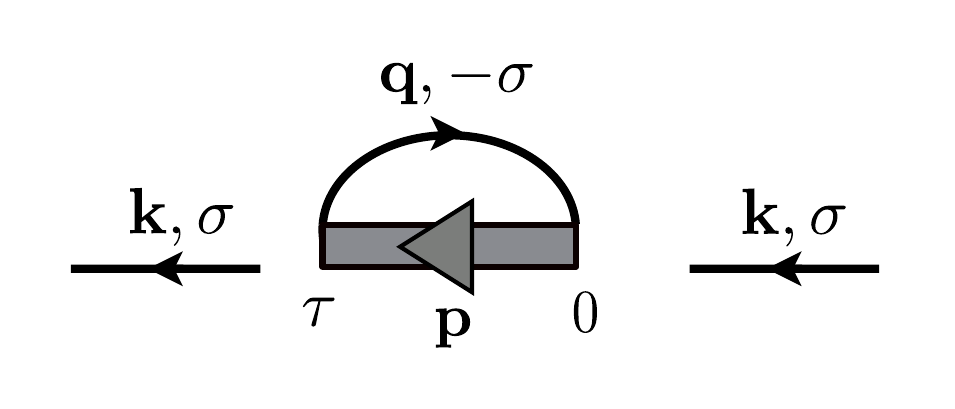}
\caption{Lowest-order bold self-energy diagram, expressing $\Sigma^{(1)}$ in terms of $G$ and $\Gamma$.
This diagram contains the dominant contributions to the self-energy at large momentum.}
\label{fig:Sigma1}
\end{figure}

To justify these statements, let us first consider the higher-order
bold diagrams for $\Sigma_\sigma(k,\tau)$.
Their contributions vanish uniformly in $\tau$ for $k\to\infty$.
Indeed, they contain internal vertices, and at some of these internal vertices, a large
momentum goes through and hence the integration over the internal time variable
is restricted to a narrow range (because $G$ and $\Gamma$ are narrow
functions of imaginary time at large momentum, cf. Sec.~\ref{sec:G_Gamma_large_k}).
We thus only need to consider the lowest-order bold self-energy diagram,
represented in Fig.~\ref{fig:Sigma1}. 
The momenta $\qq$ and $\pp$ of the
$G$ and $\Gamma$ lines are related by momentum conservation,
$\pp=\qq+\kk$.
Thus,
when $k\gg k_{\rm typ}$, at least one of the momenta $p$ and $q$ has to be $\gg k_{\rm typ}$.
\\{\it \underline{Case 1: $p\gg k_{\rm typ}$.}}
Choosing $\qq$ as the integration variable, we have $\Sigma^{(1)}_\sigma(k,\tau) = \int G_{-\sigma}(\qq,-\tau)\,\Gamma (\pp=\kk+\qq,\tau)\,d^3q/(2\pi)^3$.
As discussed in Sec.~\ref{sec:G_Gamma_large_k}, $\Gamma(p,\tau)$ is small except
in the relevant time-region $0<\tau \lesssim 1/p^2$ where it can be replaced with $\Gamma_v$.
We further observe that the relevant values of $q$ in the integral are
$\lesssim k_{\rm typ}$, an assumption that will be justified {\it a posteriori}.
This implies that $\pp\simeq \kk$,
 and thus the relevant time-region is $0<\tau\lesssim 1/k^2$.
Therefore we can replace $\Gamma (p,\tau)$ with  $\Gamma_v (k,\tau)$ and
$G_{-\sigma}(q,\tau)$ with $G_{-\sigma}(q,0^-)=n_{-\sigma}(q)$.
Since the remaining integral over $\qq$ gives us the particle density $n_{-\sigma}$,
we arrive at the result (\ref{eq:sig+},\ref{eq:def_sig+}).
Finally,
 the relevant momenta in the integral for particle density are $q \lesssim k_{\rm typ}$,
 which justifies the above assumption.
\\{\it \underline{Case 2: $q\gg k_{\rm typ}$.}}
We now choose $\pp$ as the integation variable, and write $\Sigma^{(1)}_\sigma(k,\tau) = -
\int \Gamma(\pp,\tau)\,G_{-\sigma}(\qq\,{=}\,{-}\kk{+}\pp , \beta {-} \tau) \,d^3p/(2\pi)^3$.
According to Sec.~\ref{sec:G_Gamma_large_k}, $G_{-\sigma}(q,\beta-\tau)$
is small except in the relevant time-region $0<\beta-\tau \lesssim 1/q^2$ where it can be replaced with $G_v(q,\beta-\tau)$.
We observe that the relevant values of $p$ in the integral are ${\lesssim}\,k_{\rm typ}$,
which implies that $\qq\simeq -\kk$. Thus the relevant time-region is $0<\beta-\tau\lesssim 1/k^2$,
and we can replace $G_v(q,\beta-\tau)$ with $-e^{-(k^2/2)(\beta-\tau)}$ and
$\Gamma(\pp,\tau)$ with $\Gamma(\pp,\beta^-)$. The remaining integral over $\pp$
gives us the contact, see (\ref{eq:CvsGamma}), and we readily arrive
at the result (\ref{eq:sig-},\ref{eq:def_sig-}).

\subsubsection{Tail of the momentum distribution} 

\begin{figure}
\includegraphics[width=0.7\columnwidth]{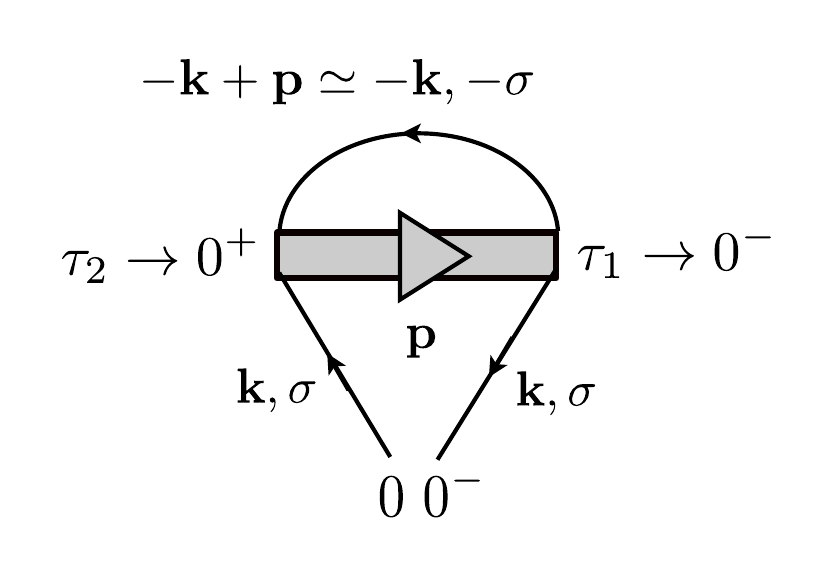}
\caption{
Leading diagrammatic contribution to the momentum distribution $n_\sigma(\kk)$ at large $k$.
The imaginary time is running from right to left.
The single-particle lines propagate forward in time 
and can be replaced with the vacuum propagators. 
The pair propagator runs backwards in time and is fully dressed.
\label{fig:diag}}
\end{figure}

In short, the tail of the momentum distribution comes from the diagram depicted in Fig.~\ref{fig:diag},
which can be interpreted physically
as the simultaneous propagation of two opposite-spin particles of large and nearly opposite momenta 
and of a missing pair  with lower momentum.
More precisely,
for $k\to\infty$ the Dyson equation
simplifies:
\bml
n_\sigma(\kk) = G_\sigma(\kk,0^-)
= G_\sigma^{(0)}(\kk,0^-)
\\
+
\int_{-\frac{\beta}{2}}^{\frac{\beta}{2}} d\tau_1 \int_{-\frac{\beta}{2}}^{\frac{\beta}{2}} d\tau_2 \, G^{(0)}_\sigma(\kk,-\tau_1) \Sigma_\sigma(\kk,\tau_1\,{-}\,\tau_2) G_\sigma(\kk,\tau_2)
\\
 \simeq  \int_{-\frac{\beta}{2}}^{\frac{\beta}{2}} d\tau_1 \int_{-\frac{\beta}{2}}^{\frac{\beta}{2}} d\tau_2 \, G_v(\kk,-\tau_1) \Sigma_\sigma(\kk,\tau_1\,{-}\,\tau_2) G_v(\kk,\tau_2).
\end{multline}
Indeed,
the ideal-gas momentum distribution decays exponentially at large $k$ so that we can neglect the term $G_\sigma^{(0)}(\kk,0^-)$,
and in the remaining term we can replace $G$ with $G_v$ according to Subsec.~\ref{sec:G_Gamma_large_k}.
Note that we took  the integration domain for the internal times $\tau_1$ and $\tau_2$ to be
$]-\beta/2;\beta/2[$
instead of the usual
$]0;\beta[$,
which is allowed since the integrand is a periodic function of $\tau_1$ and $\tau_2$.
As a result, the time-arguments of the $G_v$ factors never approach $-\beta$,
and thus the $G_v$
can be replaced by the {\it retarded} vacuum propagators,
{\it i.e.} we have $G_v(\kk,\Delta\tau) \simeq -\theta(\Delta\tau) e^{-(k^2/2)\Delta\tau}$ for $k\to\infty$ and $\Delta\tau\in]-\beta/2;\beta/2[$,
where $\theta(.)$ is the Heaviside function.
Hence the integral is dominated by $\tau_2\to0^+$ and $\tau_1\to0^-$,
and the imaginary time argument $\tau_1-\tau_2$ of the self-energy tends to $0^-$.
The asymptotic expression of $\Sigma_\sigma(\kk,\tau)$
for $k\to\infty$, $\tau\to0^-$
is known analytically, cf.~Eq.~(\ref{eq:def_sig+}).
After substitution of this expression into the asymptotic Dyson equation
given above, the large-momentum tail, Eq.~(\ref{eq:C_nk}), is recovered.

\subsection{Incorporating ultraviolet analytics into BDMC} \label{subsec:incorp}

A hallmark of BDMC is its unique capability to incorporate analytical knowledge.
The analytical considerations of the previous subsection have the following implications for our BDMC calculation.
Firstly, the contact can be evaluated accurately from the bold pair propagator thanks to the relation Eq.~(\ref{eq:CvsGamma}),
as was done in Ref.~\onlinecite{RossiContact}.
Furthermore,
since the $\Cr/k^4$ tail of the momentum distribution comes exclusively from the lowest-order self-energy diagram,
this tail is automatically built into our self-consistent BDMC scheme
 provided this  diagram is evaluated with high precision.
We achieve this by
using numerical Fourier transformations
(rather than Monte Carlo)
and analytical treatments of leading-order singularities,
in the spirit of Ref.~\onlinecite{Haussmann_PRB}, see Appendix~\ref{app:1st_order} for details.
As a result, in the BDMC data  for the momentum distribution, the $\Cr/k^4$ tail is automatically present and free of $k$-dependent noise.\cite{RossiContact}
Note that here, $\Cr$ 
comes from the fully dressed pair propagator $\Gamma$, given
by the BDMC self-consistency which includes higher-order contributions;
hence $\Cr$
differs from the one of the self-consistent
$T$-matrix approximation of  Refs.~\onlinecite{Haussmann_PRB,ZwergerViscosity}.
On the technical side, we mention that treating the lowest-order self-energy
diagram separately (without using Monte Carlo) has another advantage: the steep functions
of $\tau$ in Eqs.~(\ref{eq:def_sig+},\ref{eq:def_sig-}) would be hard to capture
by Monte Carlo sampling.

\section{Ladder scheme} \label{sec:ladder}

{\bl

As an alternative to the bold scheme discussed above,
we also employ a {\it partially} dressed scheme, in which diagrams are built from the bare single-particle propagator $G^{(0)}$ and the partially dressed pair-propagator $\Gamma^{(0)}$, defined as the sum of ladder diagrams built with $G^{(0)}$, see Eq.~(\ref{eq:Gamma0_diag}). For simplicity we will refer to this as the ``ladder scheme'' (the ladder summation being the minimal dressing procedure allowing to work with zero-range interactions in continuous space).
While the first diagrams of the ladder series for the single-particle self-energy $\Sigma$
are given by Eq.~(\ref{eq:Sig_vs_Gamma0}) above,
the ones for the pair self-energy $\Pi$ are}
\be
\fgies{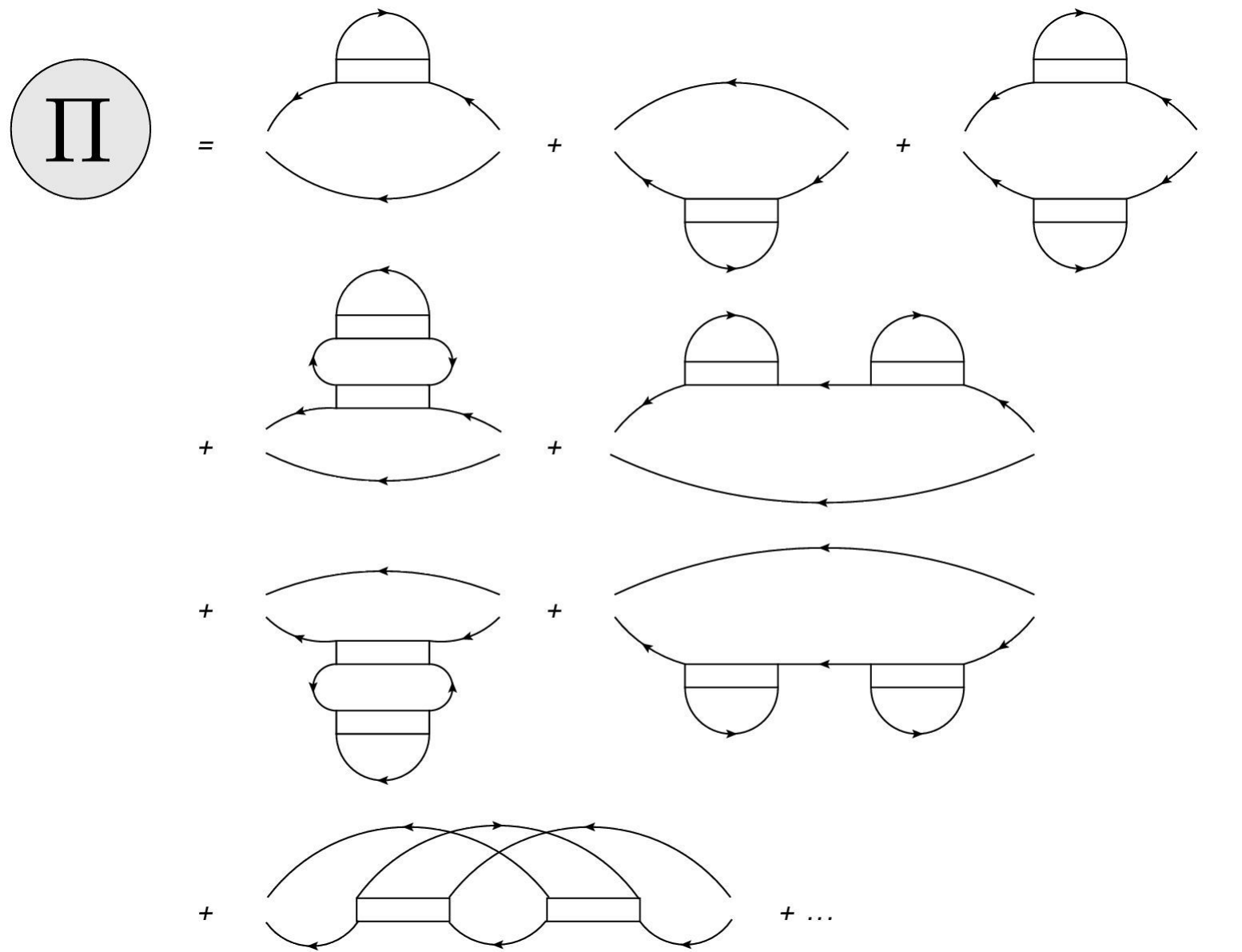}{0.3}{-1mm}
\label{eq:Pi_vs_Gamma0}
\ee

{\bl
A drawback of
the ladder scheme is that it
can only be used 
 for temperatures above
the approximate critical temperature $T_c^{(0)}$ at which $\Gamma^{(0)}(P=0,\Omega_n=0)$ diverges.
In the region $T_c^{(0)}>T>T_c$,       $\Gamma^{(0)}$ has a pole at finite momentum so that the ladder scheme cannot be used.
\footnote{For the  construction of the skeleton formalism in Sec.~\ref{sec:diag},
this is only a formal issue that does not cause any problems:
Equation~(\ref{eq:Sig_skele}) can be directly derived from
Eq.~(\ref{eq:Sigma_g0}) without using Eq.~(\ref{eq:Sig_vs_Gamma0}),
and
 only the {\it inverse} of $\Gamma^{(0)}$ appears
 in the final expressions Eqs.~(\ref{eq:dyson},\ref{eq:dyson_Gamma},\ref{eq:Sig_skele},\ref{eq:Pi_skele}).}

For the ladder scheme,
no self-consistent iterations are needed, which implies several advantages over the bold scheme:
The ladder scheme is more practical for numerical computations;
the justification of the conformal-Borel resummation method is more solid for the ladder scheme~\cite{RossiEOS,ResonLong2};
in particular,
the ladder scheme is not subject to
the misleading-convergence problems
that may potentially affect the bold scheme~\cite{Kozik_2_solutions}
(we also note that
misleading convergence was observed in Ref.~\onlinecite{Kozik_2_solutions} only for fillings near one atom per lattice site, which is a regime very different from the zero-filling limit corresponding to the present continuous-space model).

\subsection{Dyson equations}

In the ladder scheme, it is useful to consider the diagrammatic series not only for the self-energies $\Sigma$ and $\Pi$, but also for the propagators $G$ and $\Gamma$. 
In this Section, let us denote by $\Sigma^{(N)}$, $\Pi^{(N)}$, $G^{(N)}$ and $\Gamma^{(N)}$ the sum of all order-$N$  diagrams in the ladder scheme for $\Sigma$, $\Pi$, $G$ and $\Gamma$
(the number of $\Gamma^{(0)}$-lines in such diagrams is respectively
$N$, $N{-}1$, $N$ and $N{+}1$; accordingly the number of $G^{(0)}$-lines is respectively $2N{-}1$, $2N$, $2N{+}1$ and $2N$).
Note that $\Pi^{(1)}=0$
(since all $(G^{(0)}G^{(0)})$ bubbles are already contained in $\Gamma^{(0)}$).

From the Dyson equations 
\bea
G_\sigma(p,\omega_n)&=&(G^{(0)}_\sigma + G^{(0)}_\sigma \Sigma_\sigma G_\sigma)(p,\omega_n)
\label{eq:dyson_G_ladder}
\\
\Gamma(p,\Omega_n) &=& (\Gamma^{(0)} + \Gamma^{(0)} \Pi\,\Gamma)(p,\Omega_n),
\label{eq:dyson_Gamma_ladder}
\eea
we have the order-by-order Dyson equations
\bea
G^{(N)} &=& \sum_{M=1}^N G^{(0)} \Sigma^{(M)} G^{(N-M)}
\label{eq:o_by_o_dyson_G}
\\
\Gamma^{(N)} &=& \sum_{M=1}^{N} \Gamma^{(0)} \Pi^{(M)} \Gamma^{(N-M)}
\label{eq:o_by_o_dyson_Gamma}
\eea
for $1 \leq N \leq N_{\rm max}$.

As in the bold case, we need to apply a resummation procedure to extract a result from the divergent diagrammatic series. The first way to do so is to proceed exactly as in the bold case (Subsec.~\ref{sec:resum} above): apply the resummation procedure to $\Sigma$ and $\Pi$, and plug the result into the Dyson equations
(\ref{eq:dyson_G_ladder},\ref{eq:dyson_Gamma_ladder}) to get $G$ and $\Gamma$.
Another way is to apply the resummation procedure to the series $\sum_N G^{(N)}$ and $\sum_N \Gamma^{(N)}$, i.e., to use
\be
Q=Q^{(0)} + \lim_{N_{\rm max}\to\infty}\  \sum_{N=1}^{N_{\rm max}} R_N^{(N_{\rm max})}\, Q^{(N)}
\ee
 with $Q=G$ or $\Gamma$.

\subsection{Ultraviolet physics}
In the ladder scheme, the accurate incorporation of ultraviolet physics is more involved than in the bold case.

Recall that $\Sigma(p,\tau)$ and $\Pi(p,\tau)$ are narrow functions of $\tau$ when $p$ is large. This would be difficult to capture  by Monte Carlo. Our solution
for the bold code was very simple: Given that this singular behavor is completely contained (at leading order) in the
lowest-order bold diagrams, we compute these diagrams by Fourier transformation  rather than
by Monte Carlo.

For the ladder scheme, we have to do some extra work in order to achieve the same goal.
Let us denote (in the present subsection) the lowest-order bold diagrams by $\Sigma_{1, bold}$ and $\Pi_{1, bold}$.
The problem is that these bold diagrams contain an infinite number of ladder-scheme diagrams.
Our solution is as follows: During the Monte Carlo process, we do not measure the (ladder-scheme) diagrams that contribute to $\Sigma_{1, bold}$ and $\Pi_{1, bold}$. Instead, we compute them  
by combining Fourier transformation with order-by-order Dyson equations.

More precisely,
since
\bea
\Sigma_{1,bold; \sigma}(r,\tau) &=& \Gamma(r,\tau)\,G_{-\sigma}(r,-\tau)
\\
\Pi_{1, bold}(r,\tau) &=& -G_\uparrow(r,\tau)\,G_\downarrow(r,\tau),
\eea
we have 
\be
\Sigma_{1, bold; \sigma}^{(N)}(r,\tau)= \sum_{M=1}^N \Gamma^{(M-1)}(r,\tau)\,G^{(N-M)}_{-\sigma}(r,-\tau)
\label{eq:haussmann_sig_o_by_o}
\ee
for $1\leq N \leq N_{\rm max}$, and
\be
\Pi_{1, bold}^{(N)}(r,\tau) =  -\sum_{M=0}^{N-1} G_\uparrow^{(M)}(r,\tau)\,G_\downarrow^{(N-1-M)}(r,\tau)
\label{eq:haussmann_pi_o_by_o}
\ee
for $2\leq N \leq N_{\rm max}$.
Here, 
$\Sigma_{1, bold; \sigma}^{(N)}$ 
and 
$\Pi_{1, bold}^{(N)}$
denote the sum of all ladder-scheme diagrams of order $N$ that are part of the lowest-order-bold diagram. {\bl The diagrams contributing to $\Sigma_{1, bold; \sigma}^{(N)}$ up to $N=3$ are the ones in 
Eq.~(\ref{eq:Sig_vs_Gamma0}), except for the last diagram in Eq.~(\ref{eq:Sig_vs_Gamma0}) which is not part of $\Sigma_{1, bold; \sigma}^{(3)}$. 
Similarly, the diagrams contributing to $\Pi_{1, bold}^{(N)}$ up to $N=3$ are the ones in 
Eq.~(\ref{eq:Pi_vs_Gamma0}), except for the last diagram in Eq.~(\ref{eq:Pi_vs_Gamma0}) which is not part of $\Pi_{1, bold}^{(3)}$. }

We can thus perform the computations recursively in the following order:
\begin{multline}
(G^{(0)}, \Gamma^{(0)}) \longrightarrow \ldots 
\longrightarrow (G^{(N-1)}, \Gamma^{(N-1)})
\\
\longrightarrow (\Sigma^{(N)}, \Pi^{(N)})
\longrightarrow (G^{(N)}, \Gamma^{(N)})
\\\longrightarrow \ldots
\longrightarrow(G^{(N_{\rm max})}, \Gamma^{(N_{\rm max})})
\end{multline}
where at each order, the self-energies are obtained by adding up the Monte~Carlo contribution with the $(1,bold)$ contribution.

\subsection{Monte Carlo}
{\bl
The diagrammatic Monte Carlo algorithm for sampling the ladder series is
similar to the bold case described above in Sec.~\ref{sec:algo}, with the following differences. The iterative procedure (Subsec.~\ref{subsec:iter}) is not required any more.
The topologies which are reducible with respect to cutting  two {\it internal} $G_\sigma^{(0)}$ lines,
or two internal $\Gamma^{(0)}$ lines, are  sampled and measured. 
Accordingly, we perform the momentum-comparison checks described in Subsec.~\ref{subsec:ergodic} only between one internal line and the measuring line to omit one-particle reducible diagrams. 
Lastly, the $(1,bold)$ diagrams are not measured; they are identified in a way similar to detecting whether a diagram is  non-skeleton, except now we only check whether the diagram
falls apart if we cut two specific lines (the internal $G^{(0)}$-lines which are connected to the external three-point vertices). 
}

\section{Conclusion and outlook}

For spin-$1/2$ fermions with contact interactions in continuous space,
we have described a BDMC scheme allowing to sum up 
efficiently and accurately
the skeleton diagrammatic series built from single-particle propagators and pair propagators. Our procedure combines Monte Carlo sampling of higher-order diagrams with special treatment of ultraviolet singularities. We also presented an alternative ``ladder scheme'', where diagrams are built from the bare single-particle propagator and a partially dressed pair propagator; in this case the treatment of ultraviolet singularities is more involved. A crucial separate aspect of the approach is the construction of an appropriate divergent-series resummation method; this was reported in Ref.~\onlinecite{RossiEOS} 
and will be detailed elsewhere~\cite{ResonLong2}.

While the first numerical results presented in Refs.~\onlinecite{VanHouckeEOS,RossiEOS,RossiContact} 
are restricted to the unpolarized unitary gas, we expect the approach to be direcly applicable to the polarized gas throughout the BEC-BCS crossover,
as well as to the mass-imbalanced case.
Extension to two dimensions also seems feasible,
as already demonstrated for the polaron problem~\cite{Vlietinck_2D,Pollet_polaron_2D}.}
A similar scheme may be used to study the leading finite-range correction.

{\bl Another direction is the development of new algorithms to perform the summation over diagrams.
Rather than sampling stochastically topologies,
one may
sum exactly over all topologies 
at each Monte Carlo update.
With the efficient summation strategy
that was recently introduced for the Hubbard model~\cite{RossiCDet,SimkovicCDet,MoutenetCDet},
one obtains a better
computational complexity
than for
the original DiagMC~\cite{RossiComplexity}.
It can also be advantageous to perform this exact summation by brute-force enumeration provided the momentum and time variables are chosen appropriately, as sucessfully demonstrated very recently for the electron gas~\cite{KunHauleEG}.
A radically different approach would be to work with Schwinger-Dyson equations, for which new algorithms were introduced and applied to bosonic models~\cite{BuividovichSchwingerDyson,BuividovichDavody,PfefferPolletHomotopy,PfefferPolletFull}.
}

\acknowledgements
{\it To the memory of Joe~Babcock, who has played a crucial role at the UMass cluster over many years.}

We thank R.~Rossi and E.~Kozik
for a fruitful collaboration
  on the closely related Refs. \onlinecite{VanHouckeEOS,RossiEOS,RossiContact,ResonLong2}.
We are grateful to R. Haussmann, who provided us, for comparison, with unpublished propagator data obtained from the lowest-order bold diagrams as in Refs.~\onlinecite{Haussmann_PRB,HaussmannZwergerThermo}.
We thank G.~Bertsch, A. Bulgac, E. Mueller, G. Shlyapnikov and S.~Tan for  comments.

This work was supported by
the Research Foundation - Flanders FWO (K.V.H.),
ERC grants Thermodynamix and Critisup2 (F.W.),
a PICS from CNRS (F.W., N.P. and B.S.),
National Science Foundation under grant DMR-1720465, MURI Program ``Advanced quantum materials -- a new frontier for ultracold atoms''
from AFOSR
and the Simons Collaboration on the Many Electron Problem (N.P. and B.S.).
T.O. was supported
by the MEXT HPCI
Strategic Programs for Innovative Research (SPIRE),
the Computational Materials Science Initiative (CMSI)
and Creation of New Functional Devices and High-Performance Materials to Support Next Generation Industries (CDMSI), and by
a Grant-in-Aid for Scientific Research 
(No.  22104010,
 22340090, 16H06345 and 18K13477)
from MEXT, Japan.
We acknowledge the hospitality of
the Institute for Nuclear Theory, Seattle
(INT-10-1, INT-11-1).

\appendix

\section{Ladder diagrams} \label{ap:Gamma0_prop}

In this appendix, we give
some useful analytical properties of the pair propagator $\Gamma^{(0)}$ defined by
the sum of  ladder diagrams [Eq.~(\ref{eq:Gamma0_diag})] and  describe 
its numerical calculation in frequency domain.

 The expression of $\Gamma^{(0)}(\mathbf{P},\Omega_n)$ was given in Eq.~(\ref{eq:gammagen}).
For $\Omega_n \neq 0$ or $\mathbf{P}^2/4 - 2 \mu > 0$ it can be rewritten as
\bml
\frac{1}{\Gamma^{(0)}(\mathbf{P},\Omega_n)} =
\frac{1}{\td{\Gamma}^{0}(\mathbf{P},\Omega_n)}
\\+ \int  \frac{d\mathbf{k}}{(2\pi)^3}   \,   \frac{n^{(0)}_{\uparrow}(\mathbf{P}/2 + \mathbf{k}) + n^{(0)}_{\downarrow}(\mathbf{P}/2 - \mathbf{k}) }{i \Omega_n +
2 \mu - \mathbf{P}^2 /4 - \mathbf{k}^2  } \;
\label{eq:gamm4}
\end{multline}
where
\bml
\frac{1}{\tilde{\Gamma}^{(0)}(\mathbf{P},\Omega_n)} =  \frac{1}{4\pi} \left( \frac{1}{a} - \sqrt{\mathbf{P}^2 /4 -  2 \mu - i \Omega_n }\right) \;
\label{eq:Gamma0_td}
\end{multline}
and
we take the convention that the real part of the square root is positive.

In time-domain, we get (after transforming the summation over Matsubara frequencies 
into a contour integral using the residue theorem)
\bml
\tilde{\Gamma}^{(0)}(\PP,\tau) = -\frac{8}{\sqrt{\tau}}\,e^{-\left( P^2/4 - 2\mu\right) \tau}
\\
\int_0^\infty dx\ \frac{e^{-x^2}}{1-e^{-\beta \left( P^2/4 - 2\mu\right) - (\beta/\tau) x^2}},
\label{eq:Gamma_td_tau}
\end{multline}
where,
for simplicity,
 we restricted the analysis to the unitary case $a = \infty$, 
and assumed that $P^2/4 - 2\mu >0$.

{\bl
A useful property is that
in
the large-momentum short-time limit, $P\to\infty$, $\tau\to0^+$, $P^2\tau\lesssim 1$, we have
\be
\Gamma^{(0)}(P,\Omega_n) \simeq \tilde{\Gamma}^{(0)}(P,\Omega_n)\simeq \Gamma_v(P,\tau).
\ee
Indeed, in this limit, in the integrand in Eq.~(\ref{eq:Gamma_td_tau}),
the denominator tends to $1$, which  yields
$\tilde{\Gamma}^{(0)}(P,\tau) \simeq \Gamma_v(P,\tau)$,
where $\Gamma_v$ is defined in Eq.~(\ref{eq:def_Gamma_c});
moreover, in this same limit, 
we have $\Gamma^{(0)}(P,\tau) \simeq \tilde{\Gamma}^{(0)}(P,\tau) $,
because in the large-momentum large-frequency limit,
we have $\Gamma^{(0)}(P,\Omega_n) \simeq \tilde{\Gamma}^{(0)}(P,\Omega_n) $
by neglecting the Fermi factors compared to unity in
Eq.~(\ref{eq:gammagen}).

In practice we numerically compute and tabulate $\Gamma^{(0)}(\mathbf{P}, \Omega_n)$.}
We distinguish between $\Omega_n = 0$ and $\Omega_n \neq 0$.
For $\Omega_n \neq 0$ we can use the expression (\ref{eq:gamm4},\ref{eq:Gamma0_td}). The angular integration is done analytically, and one is left with a one-dimensional integral which is evaluated numerically.
When $\Omega_n=0$, we have to use the full expression Eq.~(\ref{eq:gammagen}),
whose integrand does not diverge,
because
 $2 \mu - P^2 /4 - k^2  = 0$ implies that also $1 - n_{\uparrow}^{(0)}(\mathbf{P}/2 + \mathbf{k}) - n_{\downarrow}^{(0)}(\mathbf{P}/2 - \mathbf{k}) = 0$.
The angular integration is again done analytically.

{\bl
For the ladder scheme, we also need $\Gamma^{(0)}(\mathbf{P},\tau)$,
which we obtain from $\Gamma^{(0)}(\mathbf{P}, \Omega_n)$ using the  procedure described for $\Gamma$ at the end of App.~\ref{app:dyson}.
}

\section{First order diagrams} \label{app:1st_order} 

The lowest-order diagram for the one-body and two-body self-energy is evaluated separately (without Monte Carlo), in order to accurately capture the singular behavior coming from the zero-range interaction.
Our procedure, described in detail in the following, is similar to the one of
 Ref.~\onlinecite{Haussmann_PRB},
 in that it
  uses Fourier transformation between momentum and position space,
  with analytical treatment of singular pieces.

In position space, we simply have
\be
\Sigma^{(1)}(r,\tau)=\Gamma(r,\tau)G(r,-\tau) \; .
\label{eq:Sig_Haussmann}
\ee
To Fourier transform the propagators $G$ and $\Gamma$ from momentum space to position space,
we write them as
$G=G_v+\delta G$
and
$\Gamma = \Gamma_v + \delta\Gamma$,
where $G_v$ and $\Gamma_v$
capture the leading-order large-momentum short-time behavior of $G$ and $\Gamma$,
see Eqs.~(\ref{eq:G_large_k},\ref{eq:Gamma_large_k},\ref{eq:def_G_v},\ref{eq:def_Gamma_c});
the Fourier transform of $G_v$ and $\Gamma_v$ is then done analytically while
$\delta G$ and $\delta\Gamma$ are Fourier transformed numerically.
Furthermore, to ensure that the Fourier transformation $\delta G(k\to r)$ is done accurately,
we have derived analytical expressions for the leading-order ultraviolet behavior 
of $\delta G$ both in momentum and position space, see Appendix~\ref{app:G_large_k}.

Finally, $\Sigma^{(1)}$ has to be Fourier transformed back from position to momentum space.
We again single out singular parts which we transform analytically.
We rewrite Eq.~(\ref{eq:Sig_Haussmann}) as
\bml
\Sigma^{(1)}(r,\tau)=\Gamma_v(r,\tau)G_v(r,-\tau)
+ \Gamma_v(r,\tau)\delta G(r,-\tau)
\\
+ \delta\Gamma(r,\tau) G_v(r,-\tau)
+ \delta\Gamma(r,\tau) \delta G(r,-\tau) \; .
\label{eq:firstordertrick}
\end{multline}
The  Fourier transform to momentum space is done analytically for the first term,
and numerically for the last term.
For the cross-terms (second and third term), we single out a singular piece whose Fourier transform to momentum space is done analytically:
\bml
\Gamma_v(r,\tau)\delta G(r,-\tau)=
\Gamma_v(r,\tau) \delta G(r=0,-\tau)
\\+
\Gamma_v(r,\tau)[\delta G(r,-\tau)-\delta G(r=0,-\tau)] \; .
\label{eq:gamsub}
\end{multline}
The first term in Eq.~(\ref{eq:gamsub}) is indeed singular for $\tau\to0^+$ and $r\to0$, where $\Gamma_v(r,\tau)$ becomes 
a sharply peaked function of $r$. Its Fourier transform simply gives the contribution $\Gamma_v(p,\tau) \delta G(r=0,-\tau)$ to $\Sigma^{(1)}(p,\tau)$.
The second term in Eq.~(\ref{eq:gamsub}) is Fourier transformed numerically.
The second cross-term in Eq.~(\ref{eq:firstordertrick}) is treated similarly, by writing it as
\bml
\delta\Gamma(r,\tau) G_v(r,-\tau)=
\delta\Gamma(r=0,\tau) G_v(r,-\tau)
\\+
[\delta \Gamma(r,\tau)-\delta \Gamma(r=0,\tau)] G_v(r,-\tau) \; .
\end{multline}

We note that one could think of the following alternative procedure:
 subtract the analytical singular pieces
$\Sigma^{(+)}(r,\tau)+\Sigma^{(-)}(r,\tau)$ %\; ,
from $\Sigma^{(1)}(r,\tau)$,
do the Fourier transform to momentum space, and then add back $\Sigma^{(+)}(p,\tau)+\Sigma^{(-)}(p,\tau)$.
Actually, this alternative procedure would be
essentially equivalent to the previous one, 
since
we have
\be
\Sigma^{(+)}(r,\tau)=\Gamma_v(r,\tau)  G(r=0,0^-)
\ee
\be
\Sigma^{(-)}(r,\tau)=\Gamma(r=0,\beta^-)  G_v(r,-\tau).
\ee

The first-order pair self-energy $\Pi^{(1)}$ is computed similarly, by going to position space,
the singular pieces being treated analytically.

{\bl Finally, we note that
 it is important to use an appropriate numerical treatment
of the functions and their ultraviolet singularities
(even when the leading singularities are subtracted and treated analytically).
Similarly to Ref.~\onlinecite{Haussmann_PRB},
we used non-linear grids
to tabulate the functions,
 and we computed the Fourier transforms
 using spline-interpolation and analytical evaluation of the resulting integrals.
}

\section{Dyson equations} \label{app:dyson}

To calculate the propagator $G(q,\tau)$ from $\Sigma(q,\tau)$,
we first  Fourier transform $\Sigma(q,\tau)$ to the frequency representation.
When doing so, we single out the singular parts
 $\Sigma^{(+)}(q,\tau)$ and $\Sigma^{(-)}(q,\tau)$ given in Eqs. (\ref{eq:def_sig+},\ref{eq:def_sig-}),
 whose Fourier transforms are done analytically:
\bml
\Sigma^{(+)}(q,\omega_n)=   -  4\pi \, n_{-\sigma}\, \frac{{\rm erf}\big(\sqrt{\beta}\sqrt{q^2/4-i\omega_n}  \big)}{\sqrt{q^2/4-i\omega_n}},
\label{eq:Sig_+w}
\end{multline}
\be
\Sigma^{(-)}(q,\omega_n)=  \Cr~\frac{1+e^{-\beta q^2/2}}{i\omega_n+q^2/2}.
\label{eq:Sig_-w}
\ee
This way, we take care not only of the high-momentum leading behavior of $\Sigma$,
but also of the short-time behavior of $\Sigma$  at any momentum, which is given by
\be
\Sigma_\sigma(q,\tau)\underset{\tau\to0^+}{\simeq} -4\,n_{-\sigma}\,\sqrt{\frac{\pi}{\tau}}
\ee
see Eqs.~(\ref{eq:sig+},\ref{eq:def_sig+}).

The propagator $G$ is then given in frequency representation by the Dyson equation
Eq.~(\ref{eq:dyson_G}).
When Fourier transforming this back to time representation, we treat analytically the singular piece given by $G^{(0)}$.

To calculate the dressed pair propagator $\Gamma$, we first fourier transform $\Pi(\mathbf{p},\tau)$ to the Matsubara frequency representation,  $\Pi(\mathbf{p},\Omega_n)$, and insert this into the Dyson equation Eq.~(\ref{eq:dyson_Gamma}) to obtain $\Gamma(\pp,\Omega_n)$.

Finally we need to take the Fourier transform to the time domain to get $\Gamma(\pp,\tau)$.
In order to suppress numerical errors in the form of oscillations in $\Gamma(\mathbf{P},\tau)$ as a function of $\tau$,
we treat the large-frequency short-time and large-momentum singular part analytically.
{\bl More precisely,
we write $\Gamma = \tilde{\Gamma}_v + \tilde{\delta\Gamma}$ in the momentum-time domain, where $\tilde{\Gamma}_v$ is a simple function capturing the ultraviolet behavior of $\Gamma$ whose Fourier transform to momentum-frequency domain is done analytically, while $\tilde{\delta\Gamma}$ is Fourier transformed numerically.
We take
\begin{multline}
\tilde{\Gamma}_v(P,\tau) = - 4 \sqrt{\frac{\pi}{\tau}}\,e^{-(P^2/4 - 2\mu)\tau}
\\- 4 \sqrt{\frac{\pi}{\beta}}\,e^{-\beta(P^2/4-2\mu)}
\left[
1 + \frac{1}{e^{\beta \bar{E}(P)}-1}
\right]
e^{-\bar{E}(P)\tau}
\label{eq:Gamma_tilde_v}
\end{multline}
where $\bar{E}(P) = {\rm Max}(p^2/4-2\mu,\ \ktyp^2/4)$,
whose Fourier transform to frequency domain has the analytical expression:
\begin{multline}
\tilde{\Gamma}_v(P,\Omega_n) = -4\pi\, 
\frac{{\rm erf}\left(\sqrt{(p^2/4 - 2 \mu- i \Omega_n)\beta}\right)}{\sqrt{p^2/4 - 2\mu - i \Omega_n}}
\\
+ 4 \sqrt{\frac{\pi}{\beta}} \ \frac{e^{-\beta(p^2/4-2\mu)}}{i\Omega_n - \bar{E}(P)}.
\end{multline}
In this way,
we take care of leading and higher-order singular parts of $\Gamma$ at short time and large momentum.
}

\section{Ultraviolet asymptotics
 for $G-G^{(0)}$}
\label{app:G_large_k}

In this Appendix, we %derive
give
large-momentum and short-distance asymptotic expressions  for $G-G^{(0)}$.
The derivations being rather long, we only present the final results,
which we obtained from the diagram $G^{(0)}[\Sigma^{(+)}+\Sigma^{(-)}] G^{(0)}$ where $\Sigma^{(\pm)}$ are the analytical large-momentum expressions given in Eqs.~(\ref{eq:def_sig+},\ref{eq:def_sig-}).

\subsection{Momentum space}

At large momentum, we already know that $G(q,\tau{=}\beta^-)\simeq -\Cr / q^4$.
The generalization to $\tau\in]0;\beta[$ 
is given by the following expression,
valid when $\tau$ or $\beta-\tau$ are $\lesssim1/q^2$:
\be
(G_\sigma-G^{(0)}_\sigma)(q,\tau) \underset{q\to\infty}{\simeq}{\delta G}_a(q,\tau)
\ee
where
\be
{\delta G}_a(q,\tau)=[{\delta G}^{(-)}+{\delta G}^{(+)}_A+{\delta G}^{(+)}_B+{\delta G}^{(+)}_C](q,\tau)
\ee
with
\be
\delta G^{(-)}(q,\tau) = -\frac{\Cr}{q^4} \, e^{ - \frac{q^2}{2}  (\beta-\tau)}
\label{eq:dG-}
\ee
\bml
\delta G^{(+)}_A(q,\tau)=
\frac{16\sqrt{\pi}\,n_{-\sigma}\, e^{-\frac{q^2}{4}\tau}}{q^3}
\\
\Bigg[
q\sqrt{\tau}+
i\sqrt{\pi}\left(\frac{q^2}{2}\tau + 1\right){\rm erf}\left(i\frac{q\sqrt{\tau}}{2}\right)\,e^{-\frac{q^2}{4}\tau}
\Bigg]
\label{eq:GA}
\end{multline}
\be
\delta G^{(+)}_B(q,\tau)=\Cr\,\tau\,\frac{e^{-\frac{q^2}{2}\tau}}{q^2}
\label{eq:GB}
\ee
\be
\delta G^{(+)}_C(q,\tau)=\frac{\Cr}{q^4}\,e^{-\frac{q^2}{2}\tau}.
\label{eq:GC}
\ee

\subsection{Position space}
The large-momentum behavior $\delta G_a(q,\tau)$ of $G(q,\tau)$ obtained above gives rise to a short-distance singular behavior of $G(r,\tau)$.
In order to obtain analytical expressions for this position-space behavior,
one essentially needs to take the Fourier transform of $\delta G_a(q,\tau)$ from momentum to position space.
However, this would lead to infrared divergences. To avoid this problem, we
introduce a function $\tilde{\delta G}(q,\tau)$ which has the same large-$q$ behavior than $\delta G_a(q,\tau)$
and is properly regularized at low $q$.
More precisely, we define
\be
\tilde{\delta G}(q,\tau)=[\tilde{\delta G}^{(-)}+{\delta G}^{(+)}_A+\tilde{\delta G}^{(+)}_B+\tilde{\delta G}^{(+)}_C](q,\tau)
\label{eq:4terms}
\ee
with
\be
\tilde{\delta G}^{(-)}(q,\tau)\equiv\delta G^{(-)}(q,\tau)\ \left[1-e^{-(q/q_m)^2}\right]^2
\ee
\be
\tilde{\delta G}^{(+)}_B(q,\tau)\equiv{\delta G}^{(+)}_B(q,\tau)\,\left[1-e^{-(q/q_m)^2}\right]
\ee
\be
\tilde{\delta G}^{(+)}_C(q,\tau)\equiv\delta G^{(+)}_C(q,\tau)\ \ \left[1-e^{-(q/q_m)^2} \right]^2
\ee
where
$q_m$ is a
lower momentum cutoff
whose precise value is arbitrary
 ({\it e.g.}, one can take $q_m=k_{\rm typ}$).

These four terms have the following expressions in position space:
\be
\tilde{\delta G}^{(-)}(r,\tau)
=
\frac{\Cr}{4\pi^2} r \left[
\mathcal{F}(X) -2\mathcal{F}(Y)+\mathcal{F}(Z)\right]
\label{eq:dGm_r}
\ee
where
\be
\mathcal{F}(x)=
 \mathcal{I}(x)\,\left(1+\frac{1}{x^2}\right)
 +\sqrt{\frac{\pi}{2}}\,
\frac{e^{-x^2/2}}{x} \;,
\label{eq:Fr}
\nonumber
\ee
\be
\mathcal{I}(X)
=\frac{\pi}{2}\ {\rm erf}\left(\frac{X}{\sqrt{2}}\right) \;,
\label{eq:res_I}
\nonumber
\ee
\be
X\equiv\frac{r}{\sqrt{\Delta\tau}}
,\
Y\equiv\frac{r}{\sqrt{\Delta\tau+2/q_m^2}}
,\
Z\equiv\frac{r}{\sqrt{\Delta\tau+4/q_m^2}},
\nonumber
\ee
and $\Delta\tau\equiv\beta-\tau$;
\bml
\delta G^{(+)}_A(r,\tau)=\frac{4\,n_{-\sigma}}{\pi}\,\int_0^\infty\,dx\,e^{-x^2/(2X'^2)}\,i\,{\rm erf}\left(i\frac{x}{2X'}\right)\,
\\
\times \left(
\frac{2\cos x}{x} - \frac{\sin x}{X'^2}
\right)
 \label{eq:G+A_r}
\end{multline}
with
$X'\equiv r/\sqrt{\tau}$,
\be
\tilde{\delta G}^{(+)}_B(r,\tau)=\frac{\Cr\tau}{2\pi^2r}\,\left[
\mathcal{I}(X')
-\mathcal{I}\left(\frac{1}{\sqrt{\frac{1}{X'^2}+\frac{2}{(q_m\,r)^2}}}\right)
\right],
\label{eq:3rd_term}
\ee

\be
\tilde{\delta G}^{(+)}_C(r,\tau)=-\tilde{\delta G}^{(-)}(r,\beta-\tau).
\label{eq:4th_term}
\ee

\bibliography{felix_copy}

\end{document}